\documentclass{article}
\usepackage{chemformula} 
\usepackage[T1]{fontenc} 
\usepackage{amsfonts}
\usepackage[preprint]{neurips_2023}

\author{Jingjing Zhang\\
  Department of Petroleum Engineering\\
  Texas A\&M University\\
  College Station, TX, USA\\
  \texttt{blingbling1996@tamu.edu}
  \And
  Ulisses Braga-Neto\\
    Department of Electrical \& Computer Engineering\\
  Texas A\&M University\\
  College Station, TX, USA
  \texttt{ulisses@tamu.edu} \\
  \And
Eduardo Gildin\\
Department of Petroleum Engineering\\
Texas A\&M University\\
College Station, TX, USA
\texttt{gildin@tamu.edu}}

\title{Physics-Informed Neural Networks for\\
  Multi-Phase Flow in Porous Media Considering\\
  Dual Shocks and Interphase Solubility}

\begin{document}




\maketitle

\begin{abstract}
  Physics-Informed Neural Networks (PINNs) integrate physical principles into machine learning, finding wide applications in various science and engineering fields. However, solving nonlinear hyperbolic partial differential equations (PDEs) with PINNs presents challenges due to inherent discontinuities in the solutions. This is particularly true for the Buckley-Leverett (B-L) equation, a key model for multi-phase fluid flow in porous media. In this paper, we demonstrate that PINNs, in conjunction with Welge's Construction, can achieve superior precision in handling the B-L equations in different scenarios including one shock and one rarefaction wave, two shocks connected by a rarefaction wave traveling in the same direction, and two shocks connected by a rarefaction wave traveling in opposite directions. Our approach accounts for variations in fluid mobility, fluid solubility, and gravity effects, with applications in modeling 1D water flooding, polymer flooding, gravitational flow, and CO$_2$ injection into saline aquifers. Additionally, we applied PINNs to inverse problems to estimate multiple PDE parameters from observed data, demonstrating robustness under conditions of slight scarcity and up to 5\% impurity of labeled data, as well as shortages in collocation data.
\end{abstract}


\section{Introduction}
\label{sec:intro}

Physics-Informed Machine Learning (PIML) bridges the gap between data-driven and physics-based approaches by integrating physical principles into machine learning (ML). PIML methods include training with synthetic data, post-processing to filter non-physical solutions, transfer learning, customizing neural network architectures to enforce physical constraints, and encoding governing physical laws into the loss function \citep{latrach2023critical}. PIML has various applications in subsurface energy, such as autonomous directional drilling \citep{kesireddy2023maximizing}, Pressure Transient Analysis (PTA) \citep{badawi2023physics}, geoscience data interpretation, production forecasting, reservoir characterization, and Carbon Capture, Utilization, and Storage (CCUS) simulations \citep{wang2023insights,latrach2023critical}. 

Among PIML methodologies, Physics-Informed Neural Networks (PINNs) \citep{raissi2019physics} stand out for their explicit integration of physics equations into the loss function of the neural network, steering the model toward physically plausible solutions. PINNs excel in both data-independent solving and data-driven discovery of governing equations. In forward problems, neural networks learn the solution by minimizing the losses on governing PDE residues, initial conditions, and boundary conditions. In inverse problems, PINNs use observed data to identify unknown parameters within the governing equations while adhering to the constraints imposed by PDE residues \citep{fraces2020physics}.

This study applies PINNs to solve the Buckley-Leverett (B-L) equation \citeyearpar{buckley1942mechanism} and its variants to model multi-phase fluid flow in porous media, such as water displacing oil and CO\(_2\) displacing brine. The B-L equation complexity and nonlinearity make traditional analytical or numerical solutions challenging. Previous efforts to solve B-L problems with PINNs have encountered difficulties due to solution discontinuities. Enhancement methods, such as artificial viscosity, attention-based mechanisms, and convex hull construction for the flux function, have been proposed to address these challenges. Adding an artificial viscosity term transforms the B-L equation from hyperbolic to parabolic, approximating the exact solution as the viscosity coefficient nears zero. However, this approach can lead to a smoothed shock front that mimics the diffused shock caused by truncation or discretization errors associated with numerical methods, thereby diminishing the strength of PINNs \citep{fuks2020limitations,fraces2020physics,coutinho2023physics}. Attention-based mechanisms help neural networks focus on specific data segments, adjusting the ``attention" level to different elements in the sequence, but they can complicate implementation and increase computational demands\citep{rodriguez2022physics,diab2022data}. Alternatively, Welge's Construction imposes the Rankine-Hugoniot and Oleinik entropy conditions to avoid non-physical, multi-valued solutions by constructing a convex hull for the flux function, resulting in a sharp, physically plausible shock front \citep{welge1952simplified,magzymov2021evaluation, fraces2021physics}.  

In this study, we use Welge's Construction method with vanilla PINNs to solve the original B-L equation and two challenging variants: the dual-shock B-L model accounting for gravity and the dual-shock B-L model accounting for inter-phase solubility. We examine the sensitivity of PINNs' performance in various scenarios, showing that standard PINNs handle varying mobility ratios, dip angles, and multiple discontinuities in solutions. This proficiency enables accurate prediction of engineering problems, such as water flooding, polymer flooding, mixed flow with dominant gravity, purely gravitational flow, and CO\(_2\) injection into saline aquifers. Additionally, we apply the Welge's Construction method with vanilla PINNs in inverse problems. Starting with the mobility ratio as a learnable parameter, we evaluate the impact of observed data volume and quality on PINN effectiveness, demonstrating robustness under slight data scarcity and impurity. We then conduct a two-parameter learning experiment to identify both the mobility ratio and gravity term, where our method maintained strong performance. This study underscores the essential role of Welge's Construction in tackling hyperbolic PDEs. 

The remainder of this paper is organized as follows: Section 2 provides an introduction to the training algorithm of PINNs, including basic concepts and overall workflow. Section 3 details the physical law, including the construction method of the flux function and two variants of dual-shock B-L models. Section 4 presents the results of the implementation of forward and inverse PINN training. Finally, Section 5 summarizes the key findings in the paper and discusses their implications for future research.

\section{Physics-informed neural networks}
\label{sec:PINN}
\begin{figure}
    \centering
    \includegraphics[width=1.0\linewidth]{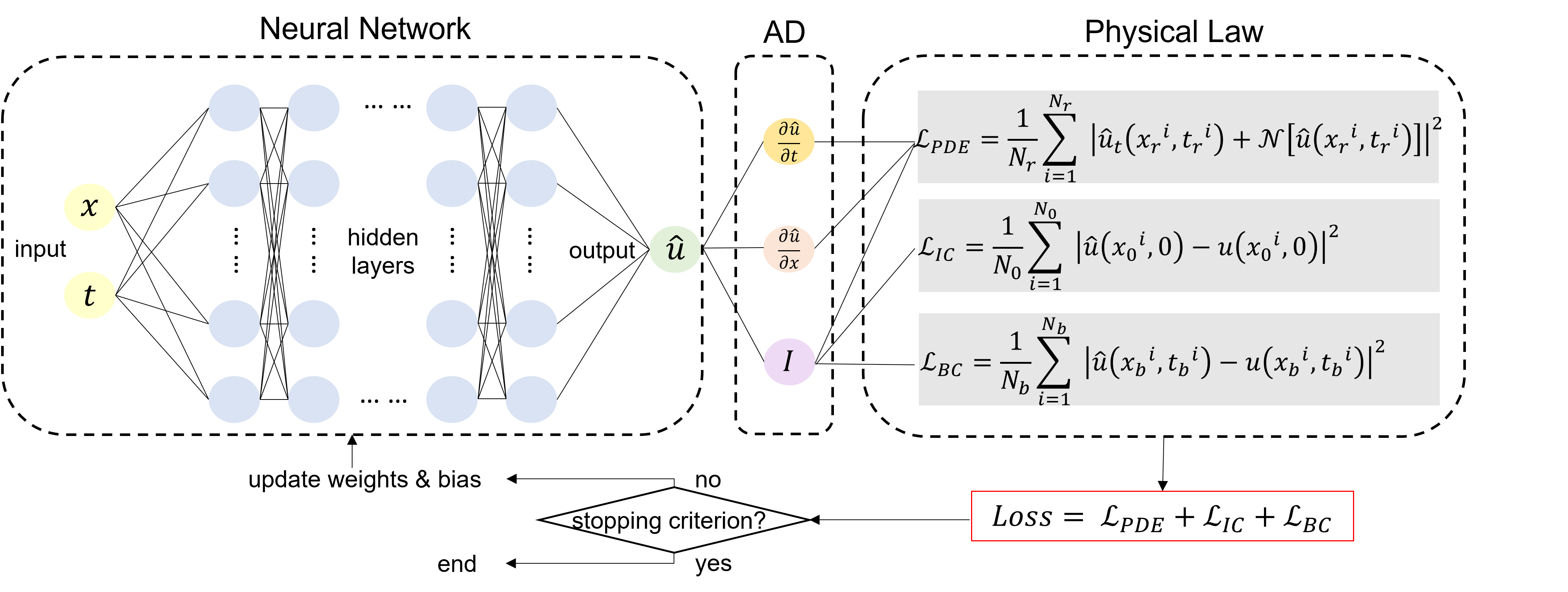}
    \caption{Forward physics-informed neural network workflow.}
    \label{fig:PINN_workflow}
\end{figure}

Physics-Informed Neural Networks (PINNs) rely on a structured workflow to model system behaviors, as illustrated in Figure.\ref{fig:PINN_workflow}. Initially, a fully connected neural network (NN) processes input data sampled across spatial and temporal domains, providing a preliminary solution estimate. Automatic differentiation (AD) then computes the derivatives of the neural network output with respect to its input coordinates and model hyperparameters. These derivatives are used to calculate the loss function and update the model parameters, respectively. The iterative process of minimizing the composite loss optimizes the model's weights and biases, refining the NN's solution $\hat{u}(x,t)$, and gradually converging towards the exact solution $u(x,t)$. 
Neural networks consist of interconnected layers (input, hidden, and output) where each neuron processes inputs through weighted sums and activation functions, making them universal function approximators. In PINNs, the NN approximates the solution to a PDE, with AD ensuring the computation of derivatives with machine precision and computational efficiency. AD systematically applies the chain rule of calculus, allowing PINNs to bypass common issues like truncation and discretization errors. 

The physical laws governing a system are integrated into the NN’s loss function. For forward problems, the aim is to predict system behavior without labeled data, acting as unsupervised learning. The loss function includes three terms: $\mathcal{L}_{P D E}$ penalizes deviations of solutions from the PDE, while $\mathcal{L}_{I C}$ and $\mathcal{L}_{B C}$ regulate the solution to adhere to the initial conditions $u\left(x_0, 0\right)$ and boundary conditions $u\left(x_b, t_b\right)$.
\begin{equation}
\text { Loss function }=\mathcal{L}_{P D E}+\mathcal{L}_{I C}+\mathcal{L}_{B C}
\label{eq:forwardloss}
\end{equation}
\begin{equation}
\mathcal{L}_{P D E}=\frac{1}{N_r} \sum_{i=1}^{N_r}\left|\widehat{u}_t\left(x_r{ }^i, t_r{ }^i\right)+\mathcal{N}\left[\widehat{u}\left(x_r{ }^i, t_r{ }^i\right)\right]\right|^2
\label{eq:L_PDE}
\end{equation}
\begin{equation}
\mathcal{L}_{I C}=\frac{1}{N_0} \sum_{i=1}^{N_0}\left|\widehat{u}\left(x_0{ }^i, 0\right)-u\left(x_0{ }^i, 0\right)\right|^2
\label{eq:L_IC}
\end{equation}
\begin{equation}
\mathcal{L}_{B C}=\frac{1}{N_b} \sum_{i=1}^o\left|\widehat{u}\left(x_b{ }^i, t_b{ }^i\right)-u\left(x_b{ }^i, t_b{ }^i\right)\right|^2
\label{eq:L_BC}
\end{equation}
Here,  $\left\{x_0{ }^i, t_0{ }^i\right\}_{i=1}^{N_0}$ denote initial condition points, $\left\{x_b{ }^i, t_b{ }^i\right\}_{i=1}^{N_b}$ represent boundary condition points, and $\left\{x_r^i, t_r^i\right\}_{i=1}^{N_r}$ are collocation points sampled within the domain of time and space. Typically, the number of collocation points greatly exceeds the number of initial or boundary condition points.

PINNs offer a versatile framework for addressing both forward and inverse problems with minimal modification in code \citep{fraces2020physics,cuomo2022scientific}. Inverse problems focus on parameter estimation from observed data, aligning with supervised learning. The loss function for inverse problems consists of two components: 
\begin{equation}
\text { Loss function }=\mathcal{L}_{P D E}+\mathcal{L}_{\text {data }}
\label{eq:inverseloss}
\end{equation}
where $\mathcal{L}_{d a t a}$ quantifies the discrepancy between model predictions and observed data as:
\begin{equation}
\mathcal{L}_{\text {data }}=\frac{1}{N_s} \sum_{i=1}^{N_s}\left|\widehat{u}\left(x_s{ }^i, t_s{ }^i\right)-u\left(x_s{ }^i, t_s{ }^i\right)\right|^2
\label{eq:L_data}
\end{equation}
$\left\{x_s^i, t_s^i\right\}_{i=1}^{N_s}$ are observed or labeled data points.

\section{Multi-phase fluid flow model}
\label{sec:Multi-phase fluid flow model}
In PINNs, integrated physical laws are often modeled by PDEs, generally represented as: 
\begin{equation}
u_t + \mathcal{N}_x[u] = 0, \text{ where } x \in \Omega \subset \mathbb{R}^d, \, t \in [0, T]
\end{equation}
where $u(x, t)$ represents the solution to be determined, $N[\cdot]$ is a nonlinear differential operator, and $\Omega$ is a subset of $\mathbb{R}^d$. This study focuses on multi-phase fluid flow dynamics modeled by the nonlinear hyperbolic PDE known as the Buckley-Leverett (B-L) equation, which involves one shock and one rarefaction wave in its original solution. More complex variants of the B-L model were later developed to account for gravity effects and inter-phase solubility. To deepen the understanding of the physical principles involved in PINNs, this section starts with the single-shock Buckley-Leverett equation and the construction method of its flux function. We will then detail the two important variants of  Buckley-Leverett equation for CO\(_2\) injection modeling and purely gravitational flow.

\subsection{Single-shock Buckley-Leverett equation}
\begin{figure}
    \centering
    \includegraphics[width=0.3\linewidth]{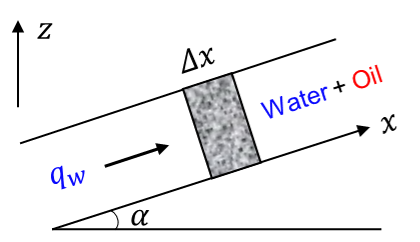}
    \caption{Schematic of 1D water flooding.}
    \label{fig:Mass_conservation}
\end{figure}

The original Buckley-Leverett equation, based on mass conservation and Darcy's law, characterizes the transport of two immiscible fluids in porous media \citep{buckley1942mechanism}:
\begin{equation}
\frac{\partial S_w}{\partial t}+\frac{\partial f_w}{\partial x}=0
\label{eq:BL6}
\end{equation}
where $S_w$ denotes the water saturation, $t$ and $x$ are dimensionless time and length, and $\alpha$ represents the angle of flow deviation from the horizontal plane, as shown in  Figure.\ref{fig:Mass_conservation}.

\begin{figure}
    \centering
    \includegraphics[width=0.5\linewidth]{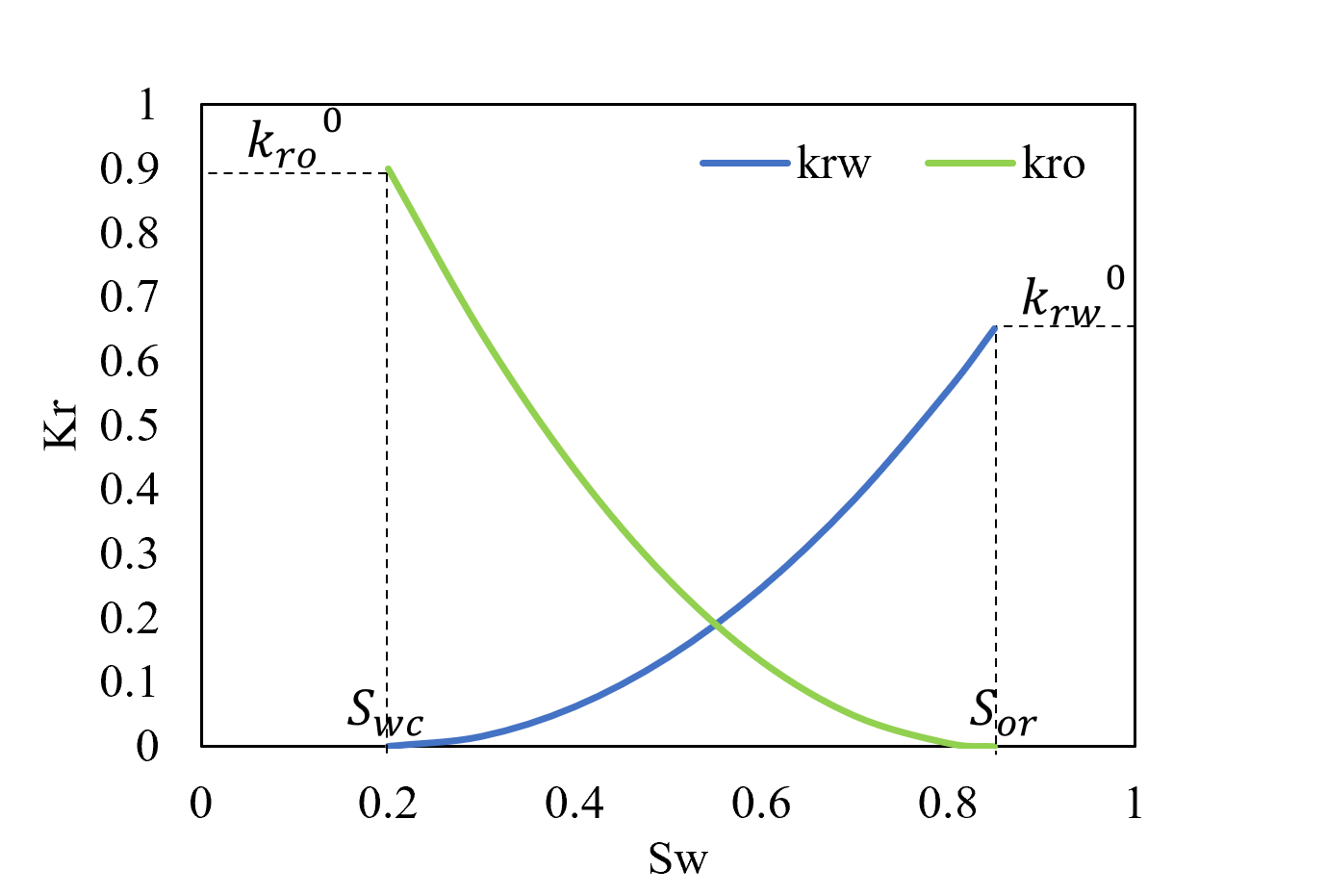}
    \caption{Relative permeability profiles for water and oil .}
    \label{fig:Rel_perm}
\end{figure}

The fractional flow or flux function $f_w$, the ratio of the water flow rate $q_w$ to the total flow rate $q_t$, is a function of water saturation as:
\begin{equation}
f_w=\frac{1-(1-S)^{n_o} N \sin \alpha}{1+\frac{(1-S)^{n_o}}{M S^{n_w}}}
\label{eq:FF10}
\end{equation}
The mobility ratio $M$ quantifies the relative mobility of two phases and the gravity number $N$ quantifies the effect of gravity on flow velocity:
\begin{equation}
M=\frac{\lambda_w}{\lambda_o}=\frac{k k_{r w}^0 / \mu_w}{k k_{r o}^0 / \mu_o}=\frac{k_{r w}^0 \mu_o}{k_{r o}^0 \mu_w}
\label{eq:FF8}
\end{equation}
\begin{equation}
N=\frac{k k_{r o}^0 A \Delta \rho g}{q_t \mu_o}
\label{eq:FF9}
\end{equation}
Here, $\mu_w$ and $\mu_o$ are the viscosities of water and oil,  $k$ is the absolute permeability of rock, $A$ is the cross-sectional area, $g$ is the gravitational constant, and $\Delta \rho=\rho_w-\rho_o$. $k_{r w}$ and $k_{r o}$  quantify the effective permeability of the medium to each fluid in the presence of both via the Corey-Brook model \citep{brooks1966properties}:
\begin{equation}
\begin{gathered}
k_{r w}=k_{r w}^0 S^{n_w} \\
k_{r o}=k_{r o}^0(1-S)^{n_0}
\end{gathered}
\label{eq:FF7}
\end{equation}
with $S$ being the effective water saturation:
\begin{equation}
S=\frac{S_w-S_{w c}}{1-S_{w c}-S_{o r}}
\label{eq:FF6}
\end{equation}
$k_{r w}^0$ and $k_{r o}^0$ are the maximum values of $k_{r w}$ and $k_{r o}$ at the endpoints of the water saturation profile ($S_{w c}$ and $1-S_{\text {or }}$ ), respectively. Figure.\ref{fig:Rel_perm} illustrates an example of $k_{r w}$ and $k_{r o}$ profiles.
The initial and boundary conditions of Eq.\ref{eq:BL6} are:
\begin{equation}
\begin{aligned}
S_w(x, t=0) &= S_{w c} \\
S_w(x=0, t) &= 1-S_{o r}
\end{aligned}
\end{equation}
The single-shock B-L model presents that when pure water is pumped into a 1D oil reservoir, at a production well ($x=1$), one obtains pure oil until the water front arrives, followed by a mixture of oil and water with increasing water cut as time goes on \citep{araujo2020numerical}. It is widely applied to model 1D horizontal water and polymer flooding processes, both important techniques to increase oil recovery in the petroleum industry.

\subsection{Construction of convex hull}
\begin{figure}
    \centering
    \includegraphics[width=0.75\linewidth]{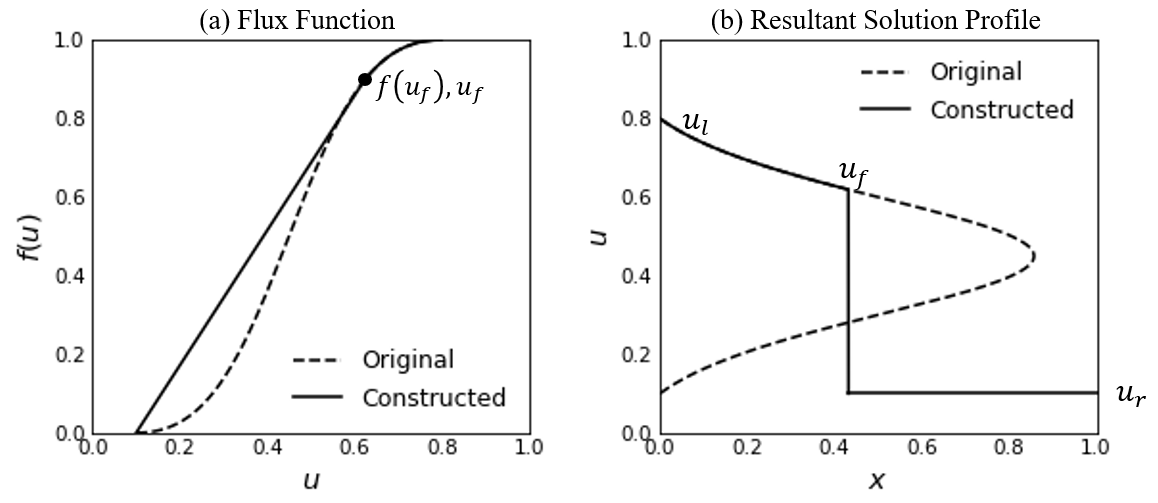}
    \caption{Example of (a) flux function  as a function of water saturation and (b) resultant saturation profile.}
    \label{fig:Flux-solution}
\end{figure}

For enhanced clarity, Eq.\ref{eq:BL6} is hereafter simplified as follows:
\begin{equation}
\frac{\partial u}{\partial t}+\frac{d f(u)}{d u} \frac{\partial u}{\partial x}=0, \quad t \epsilon[0, \infty], x \in[0,1]
\label{eq:WC1}
\end{equation}
Here, $u(x, t)$ symbolizes water saturation in water flooding scenario or gas saturation in CO\(_2\) injection scenario. A typical flux function curve is depicted with a dashed line in Figure.\ref{fig:Flux-solution}a. The derivative  $\frac{d f(u)}{d u}$ represents the propagation velocity for a specific saturation level $u$. Using this velocity to determine how far a certain saturation level travels over time enables the construction of a saturation profile, as plotted by the dashed curve in Figure.\ref{fig:Flux-solution}b. However, this profile is physically implausible because the water saturation is triple-valued at a single location. This irrationality originates from the non-convex nature of the flux function, where the velocity $\frac{d f(u)}{d u}$ initially increases with increasing saturation, peaks, and subsequently decreases, causing higher saturation levels to overtake lower ones and form a shock front \citep{dake1983fundamentals}. Because of the shock, the mathematical expression of the Buckley-Leverett problem in Eq.\ref{eq:FF10}, assuming continuity and differentiability of $u$, fails to accurately represent the dynamics ahead of the shock, rendering it inadequate for integration in PINNs as the governing equation. 

To develop a fully valid solution for the Buckley-Leverett problem, we introduce the concept of Riemann problems, which are hyperbolic conservation laws accompanied by piecewise initial conditions:
\begin{equation}
u(x, 0)= \begin{cases}u_l, & x \leq 0 \\ u_r, & x>0\end{cases}
\label{eq:WC2}
\end{equation}
The Buckley-Leverett model with its non-convex flux function is a classic example of a Riemann problem, where the left state is $u_l=1-S_{o r}$ and the right state is $u_r=S_{w c}$. In numerical analysis, the imposition of Rankine-Hugoniot jump condition and the E-condition of Oleinik on the flux function helps to select the unique and proper solution for Riemann problems \citep{leveque1992numerical}. The E-condition of Oleinik, as shown in Eq.\ref{eq:WC3}, ensures that the rarefaction wave trailing the shock does not surpass it, thereby adhering to the second law of thermodynamics, which mandates that entropy in an isolated system should not decrease: 
\begin{equation}
\frac{f(u)-f\left(u_r\right)}{u-u_r} \leq \frac{d f\left(u_f\right)}{d u_f} \leq \frac{f(u)-f\left(u_l\right)}{u-u_l}
\label{eq:WC3}
\end{equation}
The Rankine-Hugoniot jump condition equates the flow rates of the displacing fluid on either side of the shock, ensuring the conservation of mass across the shock front. It helps determine the speed at which a shock wave moves through the medium, as shown in Eq.\ref{eq:WC4}. 
\begin{equation}
\frac{d f\left(u_f\right)}{d u_f}=\frac{f\left(u_l\right)-f\left(u_r\right)}{u_l-u_r}
\label{eq:WC4}
\end{equation}
So, from points at$\left(u_r, 0\right)$ to $\left(u_f, f\left(u_f\right)\right)$, the original $f(u)$ curve is replaced by a straight line segment to represents a shock jumping from $u=u_r$ to $u=u_f$, forming a convex hull. Behind the shock, in the saturation range of $\left[u_f, u_l\right]$, Eq.\ref{eq:FF10} remains valid \citep{welge1952simplified}. With this construction, the solution can be obtained as shown by the solid line in Figure.\ref{fig:Flux-solution}b. 

\subsection{Dual-shock B-L model with inter-phase solubility}
\begin{figure}
    \centering
    \includegraphics[width=0.55\linewidth]{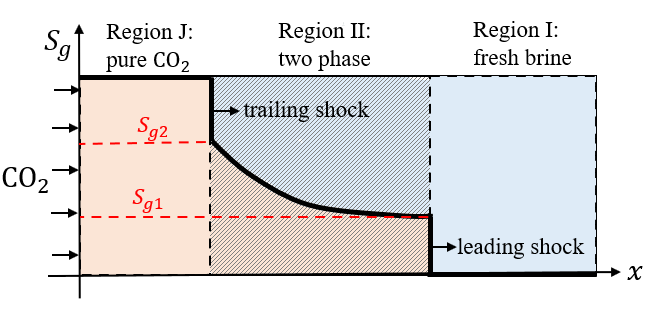}
    \caption{Schematic of a miscible gas-water displacement. Two saturation shocks divide the medium into three regions.}
    \label{fig:CO2injection}
\end{figure}

Injecting CO\(_2\) into subsurface saline aquifers can reduce greenhouse gas emissions and combat climate change \citep{green1998enhanced,li2024investigation}. The original Buckley-Leverett model, assuming strict immiscibility between phases, was modified to include a retardation factor that accounts for the partial solubility of CO\(_2\) and brine \citep{noh2007implications,burton2009co2,azizi2013approximate,bai2024storage}. This adjustment creates a dynamic two-phase, two-component system with constant phase properties. The system  involves three distinct regions: pure CO\(_2\) region (\text{I}), fresh brine region (\text{II}), and two-phase region (J) where CO\(_2\) and H\(_2\)O dissolve in each other's phases, indicating hydrodynamic and solubility trapping of CO\(_2\). These three regions are separated by leading and trailing shocks with saturation $S_{g 1}$ and $S_{g 2}$, as illustrated in Figure.\ref{fig:CO2injection}. 

\begin{figure}
    \centering
    \includegraphics[width=0.5\linewidth]{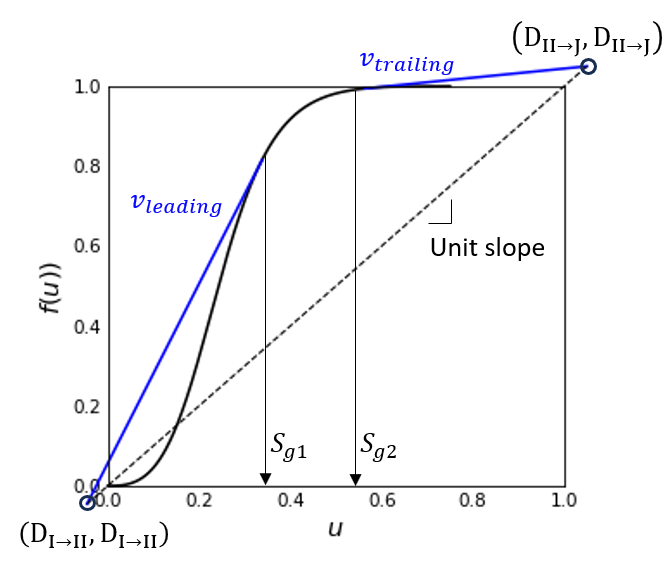}
    \caption{Construction of the flux function for miscible gas-water displacement.}
    \label{fig:Flux_function_gas}
\end{figure}

The velocities of the leading and trailing shocks depend on the solubilities of CO\(_2\) in the aqueous phase and H\(_2\)O in the gaseous phase, respectively. Figure.\ref{fig:Flux_function_gas} demonstrates the construction of the flux function by drawing tangent lines from points $(\mathrm{D}_{\mathrm{I} \rightarrow \mathrm{II}},\mathrm{D}_{\mathrm{I} \rightarrow \mathrm{II}})$ and  $(\mathrm{D}_{\mathrm{II} \rightarrow \mathrm{J}},\mathrm{D}_{\mathrm{II} \rightarrow \mathrm{J}})$to intersect the original flux function at $S_{g 1}$ and $S_{g 2}$.  The slopes of these lines dictate the velocities of the shocks, as detailed in Eq.\ref{eq:Gas1} and Eq.\ref{eq:Gas2}.

\begin{equation}
v_{\text {leading }}=\left.\frac{d f(u)}{d u}\right|_{S_{g 1}}=\frac{f\left(S_{g 1}\right)-D_{\mathrm{I} \rightarrow \mathrm{II}}}{S_{g 1}-D_{\mathrm{I} \rightarrow \mathrm{II}}}
\label{eq:Gas1}
\end{equation}
\begin{equation}
v_{\text {trailing }}=\left.\frac{d f(u)}{d u}\right|_{S_{g 2}}=\frac{f\left(S_{g 2}\right)-D_{\mathrm{II} \rightarrow J}}{S_{g 2}-D_{\mathrm{II} \rightarrow J}}
\label{eq:Gas2}
\end{equation}
The retardation factors $D$, which quantify inter-phase mass transfer (mutual solubility) as following equations, are influenced by temperature, pressure, and salinity, shaping the interactions within each phase. 
\begin{equation}
\mathrm{D}_{\mathrm{I} \rightarrow \mathrm{II}}=\frac{C_{C O_2, a}^{\mathrm{II}}}{C_{\mathrm{CO}_2, a}^{\mathrm{II}}-C_{C O_2, g}^{\mathrm{II}}}
\label{eq:Gas3}
\end{equation}
\begin{equation}
D_{\mathrm{II} \rightarrow J}=\frac{C_{\mathrm{CO}_2, a}^{\mathrm{II}}-C_{\mathrm{CO}_2, g}^J}{C_{\mathrm{CO}_2, a}^{\mathrm{II}}-C_{\mathrm{CO}_2, g}^{\mathrm{II}}}
\label{eq:Gas4}
\end{equation}
$\mathrm{D}_{\mathrm{I} \rightarrow \mathrm{II}}$ and $\mathrm{D}_{\mathrm{II} \rightarrow \mathrm{J}}$, located on the line with a unit slope through the origin, correspond to conditions in pure brine (initial) and pure CO\(_2\) (injection), respectively. Typically, CO\(_2\)'s solubility in water exceeds H\(_2\)O's solubility in gas, causing the leading shock to advance faster than the trailing shock. The trailing shock disappears when the slope of $v_{\text {trailing }}$ reaches zero, corresponding to zero water solubility in gas, with $s_{g 2}=1-s_{w r}$. 

\subsection{Dual-shock B-L model with dominant gravity}
\begin{figure}
    \centering
    \includegraphics[width=0.4\linewidth]{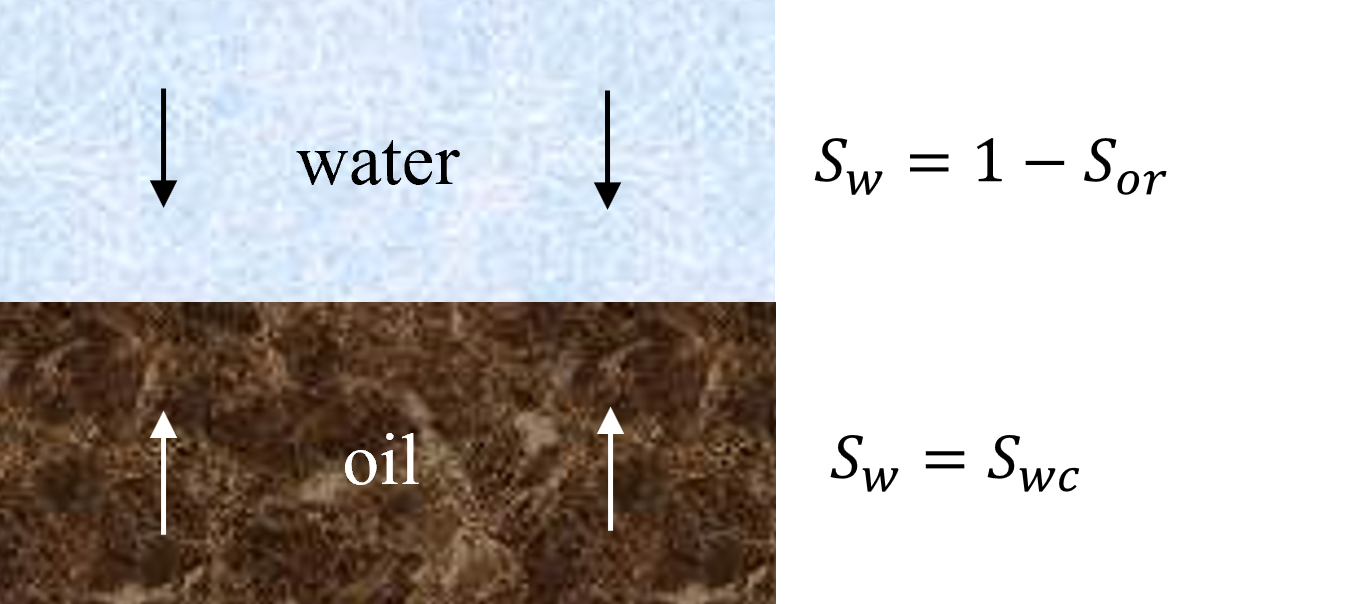}
    \caption{Initial saturation distribution of purely gravitational flow.}
    \label{fig:gravity_IC}
\end{figure}

The non-zero gravity term $N \sin\alpha$ in Eq.\ref{eq:FF10} signals that the horizontal  multi-phase flow is subject to gravity effect. However, when gravity completely dominates the flow, the fluid dynamics are described by Buckley-Leverett equation with a different flux function \citep{araujo2020numerical}: 
\begin{equation}
f\left(s_w\right)=\frac{s_w^2}{s_w^2+\frac{\mu_w}{\mu_o}\left(1-s_w\right)^2}\left[\left(1-s_w\right)^2 \frac{\mu_w}{\mu_o}\left(1-\frac{\rho_o}{\rho_w}\right)\right]
\label{eq:f_gravity}
\end{equation}
This setup assumes $n_w=n_w=2$ and $S_{w_c}=S_{o_r}=0$. A typical initial condition of such purely gravitational flow is displayed in Figure.\ref{fig:gravity_IC}. Due density difference, lighter fluid moves upwards and heavier fluid moves downwards, which is also referred to as counter-current flow. 

\begin{figure}
    \centering
    \includegraphics[width=0.5\linewidth]{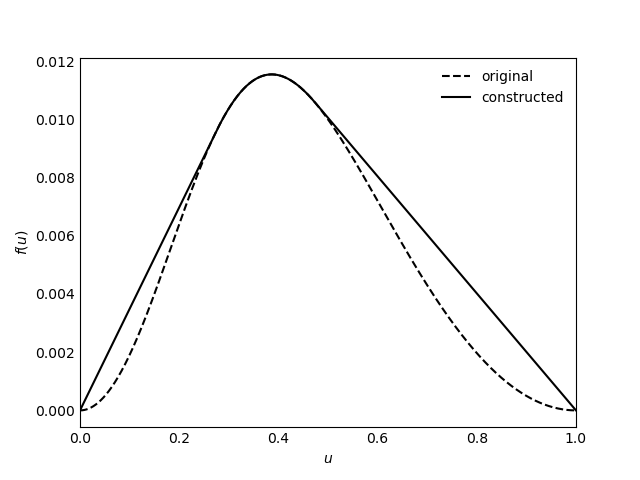}
    \caption{Example of flux function of purely gravitational flow.}
    \label{fig:Flux_gravity}
\end{figure}

The flux function exhibits multiple inflection points, as denoted by the dashed line in Figure.\ref{fig:Flux_gravity}. According the entropy and jump conditions, the construction of the bell-shaped flux function is displayed by the solid curve: from $u_r$ to the low-saturation shock and from high-saturation shock to $u_l$, the original curves are replaced by convex hulls. It can be observed that the velocity of low saturations is positive while the velocity of high saturations is negative, leading to two front shocks traveling in opposite directions.

\section{Implementation and training results}
\label{sec:Implementation and results}
This section presents the results of applying the PINN framework to both forward and inverse problems. The neural network structure utilized here was simple and straightforward, featuring an input layer with two neurons (for spatial and temporal inputs), eight hidden layers with 20 neurons each, and a single-neuron output layer (for the solution). The entire implementation was carried out with TensorFlow 2.x, and we employed the \verb|keras.models.Sequential()| method for PINN model development, initializing hyperparameters with the Xavier method and optimizing with the Adam optimizer. \verb|tf.GradientTape()| was used to compute gradients of the solution with respect to time and space for PDE residues, as well as gradients of the loss concerning model hyperparameters for PINN training. The typical runtime for a case on a T4 GPU system in Google Colab ranged from 10 to 20 minutes.

\subsection{Forward problems}
The objective of forward training was to solve the Buckley-Leverett equation by minimizing a composite loss function that included residual loss, initial condition loss, and boundary condition loss. The training dataset consisted of 10,000 collocation points within the solution domain, along with 300 data points to enforce the initial condition and another 300 data points to enforce the boundary condition, all generated using the Latin Hypercube Sampling (LHS) method. No labeled data was used for this process. The maximum iteration number was set at 20,000. 

\begin{figure}
    \centering
    \includegraphics[width=0.95\linewidth]{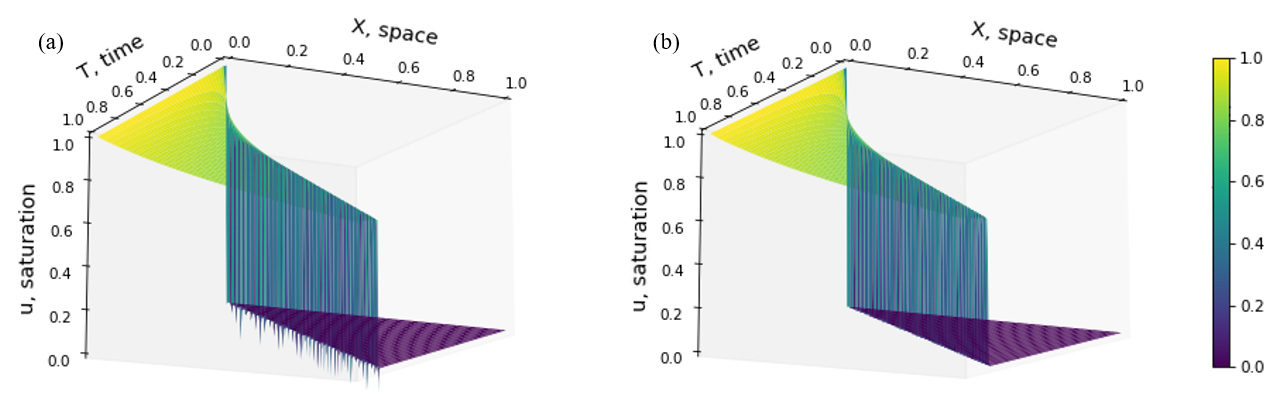}
    \caption{3D visualization of Buckley-Leverett solutions with (a) tanh and (b) sigmoid as the activation function for the output layer.}
    \label{fig:Sigmoid_necessity}
\end{figure}

Initially, the hyperbolic tangent (tanh) function was chosen as the activation function across all layers. This setup, however, led to non-physical results where the solution values fell below zero at the shock front, contrary to the expectation that water saturation levels should remain within the [0, 1] interval. To remedy this, we transitioned to using the sigmoid function for the output layer's activation, ensuring solution values were constrained within the appropriate range. This change led to a more accurate solution map, as displayed in Figure.\ref{fig:Sigmoid_necessity}.

The subsequent subsections will explore the results from forward training for the single-shock scenario, examine the sensitivity of PINN performance to various fluid mobility, incorporate gravity into the governing equation, and subsequently tackle two dual-shock Buckley-Leverett models.

\subsubsection{Base case}
\begin{figure}
    \centering
    \includegraphics[width=0.8\linewidth]{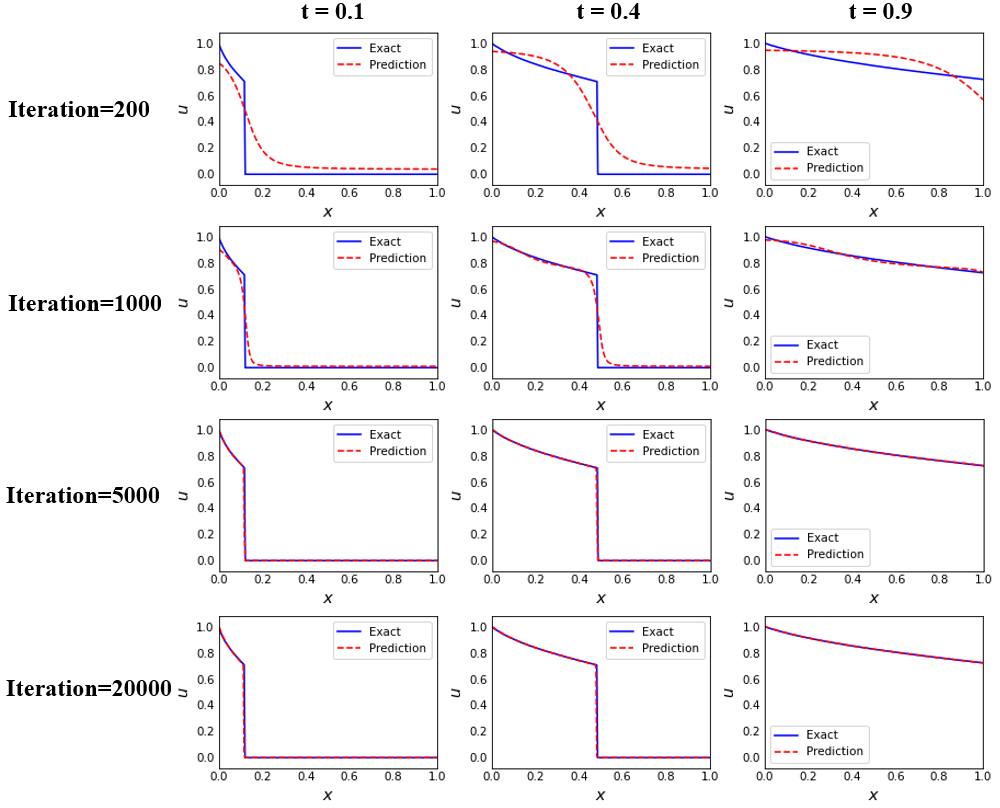}
    \caption{Evolution of solution profiles during forward PINN training for the base case.}
    \label{fig:Base_case}
\end{figure}

In the base case, we excluded the influence of gravity and use a unit mobility ratio ($M$).  Osher's method was utilized to compute analytical solutions, serving as benchmarks for evaluating the PINN predictions\citep{ketcheson2020riemann}. The progression of the training is illustrated in Figure.\ref{fig:Base_case}, displaying how the solution's profile changed over distance at specific time intervals (0.1, 0.4, and 0.9) through various stages of iteration (200, 1000, 5000, and 20000). By approximately the 5000th iteration, the trained solution closely aligned with the exact solution. The ultimate $L_2$-norm error of the PINN solution was 4.55\% and an  $L_2$-norm loss is calculated at 1.36E-6, according to Eq.\ref{eq:forwardloss}.
We explored four additional cases to assess the flexibility of standard PINNs. The details of these cases, including their specific parameters, losses, and errors, were compiled in Table\ref{tab:forward_cases}. These cases were categorized into two groups to facilitate sensitivity analyses. 
\begin{table}
  \caption{Summary of Single-shock Forward PINN Training Cases}
  \label{tab:forward_cases}
  \centering
  \begin{tabular}{lcccc}
    \hline
    \textbf{Case} & \textbf{M} & \textbf{Gravity Term} & \textbf{Error} & \textbf{Loss} \\
    \hline
    base &  1   & 0  & 0.04546  & 1.36E-06 \\
    L1   & 0.1  & 0  & 0.01408  & 7.18E-06 \\
    L2   & 10   & 0  & 0.02164  & 4.86E-06 \\
    N1   & 1    & -3 & 0.01229  & 1.19E-06 \\
    N2   & 1    & 3  & 0.01243  & 2.59E-06 \\
    \hline
  \end{tabular}
\end{table}

\subsubsection{Sensitivity analysis on fluid mobility }
The first group of cases (base, L1, L2) investigated the impact of the mobility ratio. In the base case, a unity $M$ reflects equal mobility between the displacing (water) and displaced (oil) phases. For water flooding, $M$ is typically less than 1 due to water being more mobile than oil. In contrast, for polymer flooding, where the viscosity of the displacing phase is intentionally increased through polymer addition, $M$ can be significantly higher. Extreme values were examined by setting $M=0.1$ for the L1 case and $M=10$ for the L2 case. 

\begin{figure}
    \centering
    \includegraphics[width=0.7\linewidth]{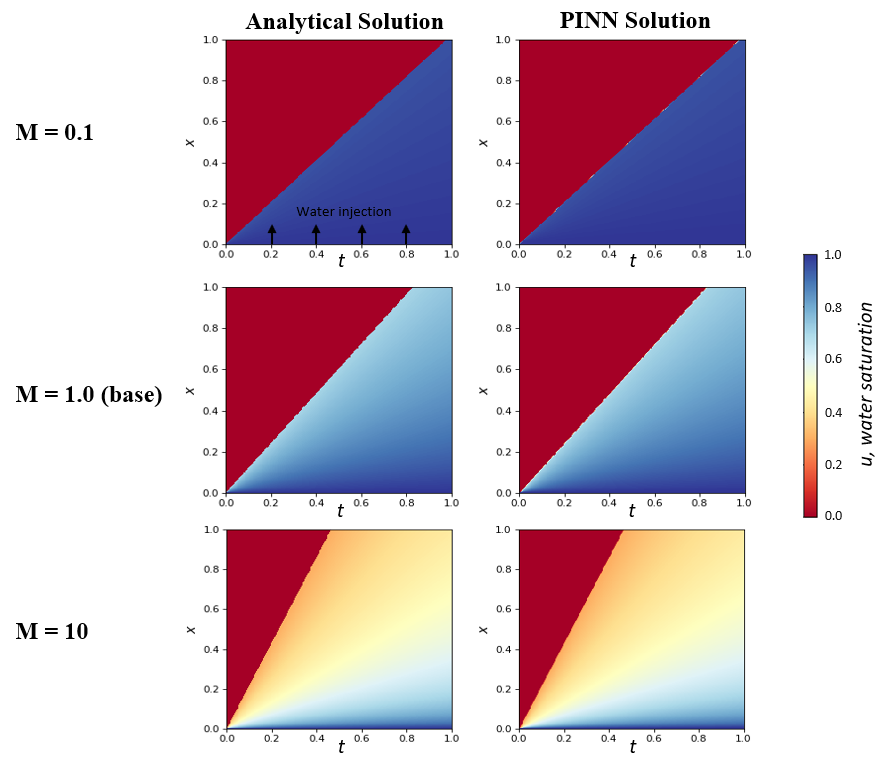}
    \caption{Analytical vs. PINN solution profiles: comparative 2D views for cases with M=0.1(first row), M=1 (second row), and M=10 (third row).}
    \label{fig:2D_M}
\end{figure}

\begin{figure}
    \centering
    \includegraphics[width=0.85\linewidth]{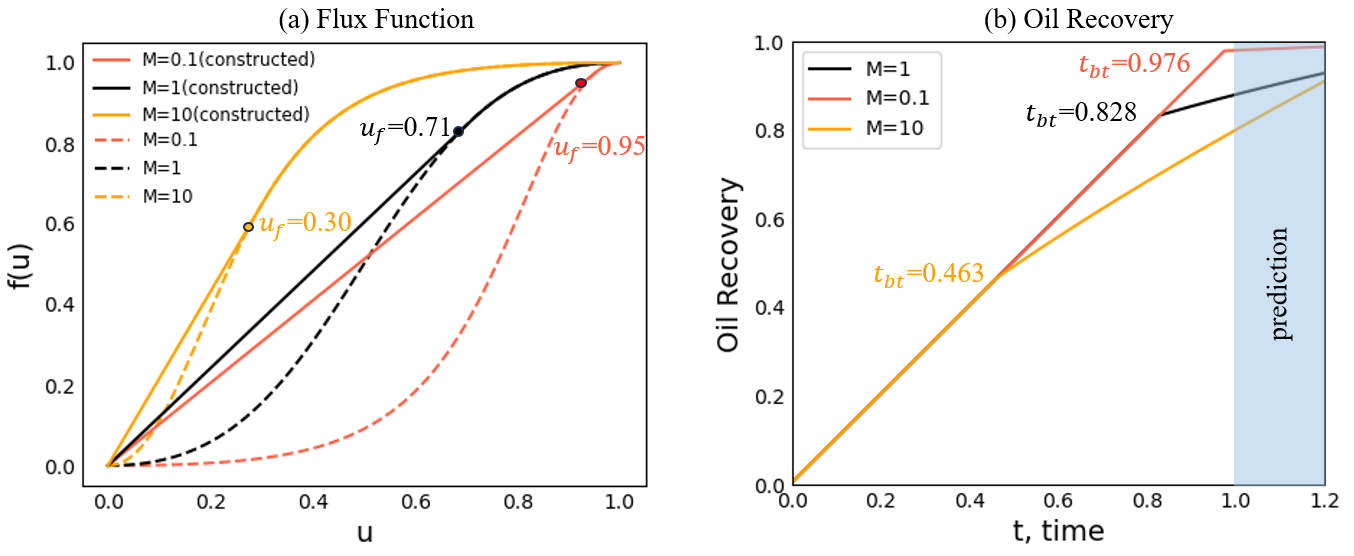}
    \caption{(a) Fractional flow curves and (b) oil recoveries of cases with M=0.1, 1, and 10.}
    \label{fig:Sensitivity_M}
\end{figure}

As summarized in Table\ref{tab:forward_cases}, cases L1 and L2 yielded comparably low errors and losses relative to the base case, indicating the proficiency of standard PINNs in resolving Buckley-Leverett models across different $M$ values. This was further evidenced by a side-by-side 2D comparison of analytical and PINN solutions in Figure.\ref{fig:2D_M}, displaying cooler hues for higher water saturation and warmer ones for higher oil saturation. There was no noticeable difference between the PINN and analytical solutions.  When comparing cases with different $M$, a smaller $M$ led to a higher front saturation $u_f$ and a delayed breakthrough time of water $t_{b t}$ (the value of $t$ when $x=1$). The shock front is marked by the transition between cold and warm color zones. 
The influence of $M$ on flux function and oil recovery was further illustrated in Figure.\ref{fig:Sensitivity_M}. On the left, the points where the dashed line (original fractional flow) intersected with the solid line (constructed fractional flow) signified $u_f$, with values of 0.30, 0.71, and 0.95 for the three cases, respectively. 
On the right, oil recovery factors were calculated by integrating the oil rate ($1-u$) over time at the producer's location ($x=1$), with the slope representing oil production rate. Oil was produced at a constant rate across all cases until the water breakthrough occurred at different times: 0.463 for the low-mobility-ratio case, 0.828 for the base case, and 0.976 for the high-mobility-ratio case. Upon breakthrough, the water cut at the producer jumped from 0 to $u_f$ and continued to rise as flooding progresses through the reservoir. A high mobility ratio led to an early breakthrough, leaving significant oil unrecovered—an unfavorable scenario. A moderate mobility ratio delayed the breakthrough and improved sweep efficiency. A low mobility ratio caused a late breakthrough, enabling nearly complete oil recovery, making it the most advantageous scenario.

Additionally, Figure.\ref{fig:Sensitivity_M}b demonstrated oil recovery predictions until a dimensionless time of 1.2, beyond the training data's range of [0,1]. The blue region visually represents the remarkable extrapolation ability of PINNs to forecast outcomes once the model has effectively learned the underlying physics. 

\subsubsection{Sensitivity analysis on gravity}
The second set of cases (base, N1, N2) incorporated the gravity term ($N \sin \alpha$) into the Buckley-Leverett equation, adding complexity to the flux function. The gravity term, as defined in Eq.\ref{eq:FF9}, is influenced by the reservoir's inclination angle $\alpha$, rock permeability, the density difference between water and oil, and the flow rate. A positive $\alpha$ indicates upward flooding, whereas a negative $\alpha$ suggests downward flooding. We explored $N \sin \alpha$ values ranging from -3 to 3. 

\begin{figure}
    \centering
    \includegraphics[width=0.7\linewidth]{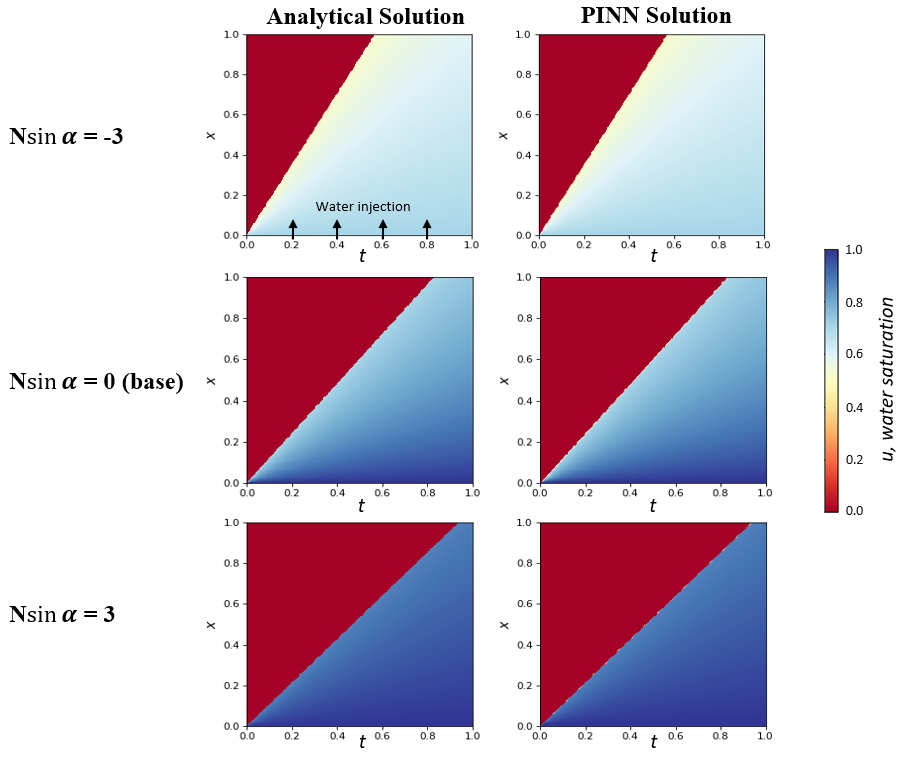}
    \caption{Analytical vs. PINN solution profiles: 2D comparison for  cases with $N\sin\alpha$=-3 (first row), $N\sin\alpha$=0 (second row), and $N\sin\alpha$=3 (third row).}
    \label{fig:2D_N}
\end{figure}

PINN training results from the N1 and N2 cases, detailed in Table\ref{tab:forward_cases}, showed small errors and losses, demonstrating the effectiveness of PINNs in solving gravity-modified Buckley-Leverett models. Figure. \ref{fig:2D_N} reinforced this point through a 2D comparison, affirming PINNs' precision in capturing the comprehensive solution landscape for these scenarios. A negative gravity term resulted in faster water breakthrough and reduced front saturation, compared to a positive one. 

\begin{figure}
    \centering
    \includegraphics[width=0.85\linewidth]{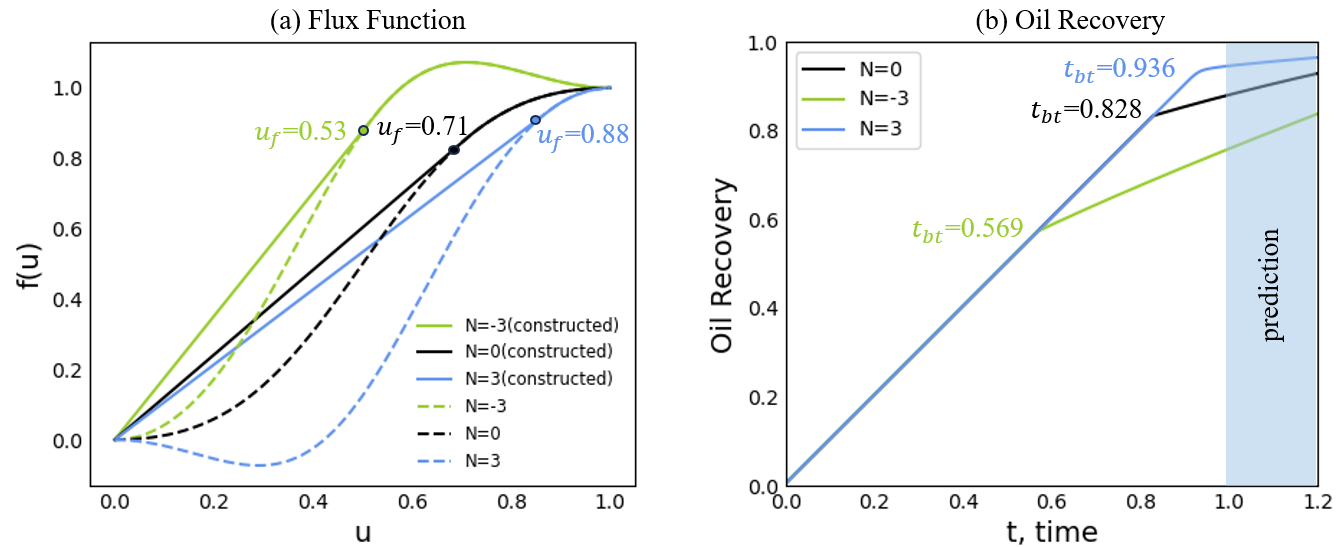}
    \caption{(a) Fractional flow curves and (b) oil recoveries of cases with $M$=-3, 0, and 3.}
    \label{fig:Sensitivity_N}
\end{figure}

Furthermore, the negative gravity effect modifies the initial conditions of the displacement process. Due to the density difference between oil and water, saturation distribution changes immediately after water injection begins. As a result, modeling of down-dip flooding cases requires careful adjustment of new boundary conditions from the original value of  $1-S_{o r}$.
Figure.\ref{fig:Sensitivity_N} examined gravity's impact on the fractional flow curve and oil recovery. In scenarios with steeply downward-dipping reservoirs, the value of  $f(u)$ may exceed one, promoting water flow while restricting oil production. Conversely, up-dip flooding impairs water flow, resulting in high front saturation $u_f$ and slow movement. Thus, a positive gravity term leads to more oil displacement at breakthrough, albeit occurring later, as shown in Figure.\ref{fig:Sensitivity_N}b. A late $t_{b t}$ is more favorable because displacement efficiency tends to be poor after breakthrough. Predictive modeling extended the system's future solution behaviors until $t$=1.2. 
Until now, vanilla PINNs have managed to model the Buckley-Leverett equation featuring a single shock in the solution. Yet, whether PINNs can effectively tackle scenarios with two discontinuities in the governing equation remains to be seen. To investigate this, PINNs are further trained on adapted Buckley-Leverett equations that model gas-displacing-water processes and purely gravitational flows.

\subsubsection{Dual-shock B-L model for CO\(_2\) injection}
Move on to the dual-shock B-L model, the data we used was sourced from Noh et al. \citep{noh2007implications}, including retardation factors of -0.45 ($\mathrm{D}_{\mathrm{I} \rightarrow \mathrm{II}}$ ) and 1.05 ($\mathrm{D}_{\mathrm{II} \rightarrow \mathrm{J}}$) to account for the solubility of CO\(_2\) in the aqueous phase and H\(_2\)O component in the gaseous phase. The viscosities for reservoir brine and injected CO\(_2\)  were set at 0.548 cp and 0.189 cp, respectively, with connate water saturation at 0.25 and residual gas saturation at 0. The reservoir conditions were maintained at 50 \(^\circ\text{C}\) and 5000 kPa. The flux function was constructed as shown in Figure.\ref{fig:Flux_function_gas}.

\begin{figure}
    \centering
    \includegraphics[width=0.8\linewidth]{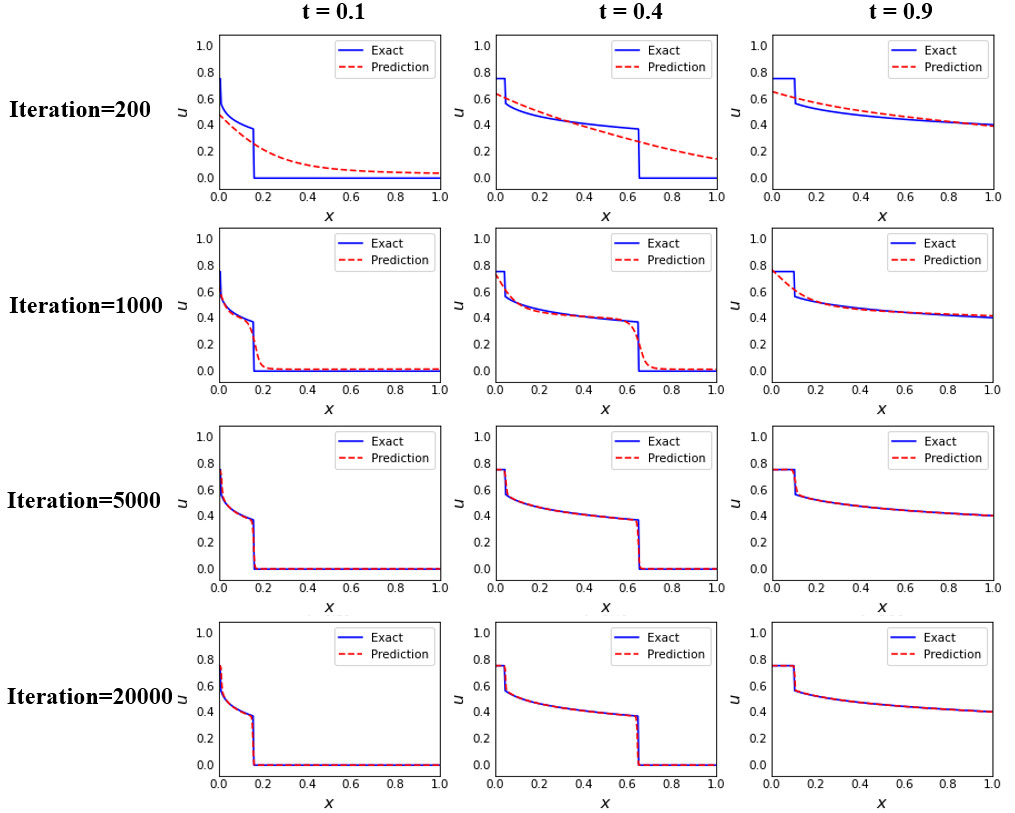}
    \caption{Evolution of PINN solution for the model with two shocks traveling in the same direction.}
    \label{fig:Gas_case}
\end{figure}

\begin{figure}
    \centering
    \includegraphics[width=0.65\linewidth]{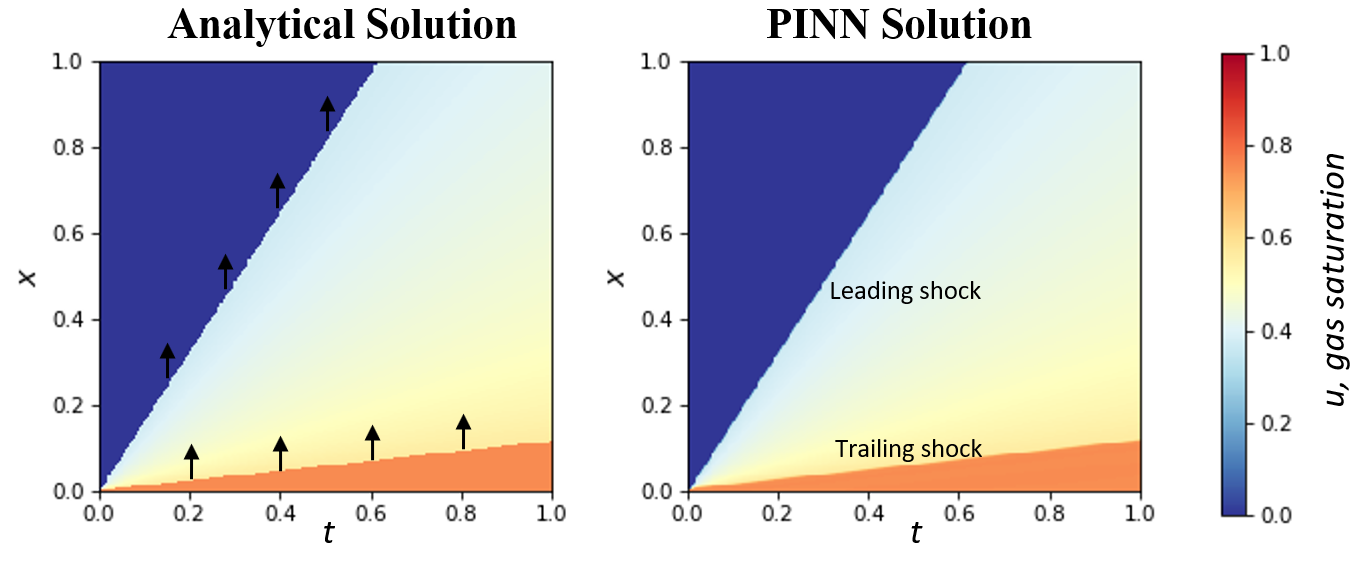}
    \caption{Analytical vs. PINN solution profiles: 2D comparison for the dual-shock B-L model for CO\(_2\) injection.}
    \label{fig:2D_gas}
\end{figure}

\begin{figure}
    \centering
    \includegraphics[width=0.8\linewidth]{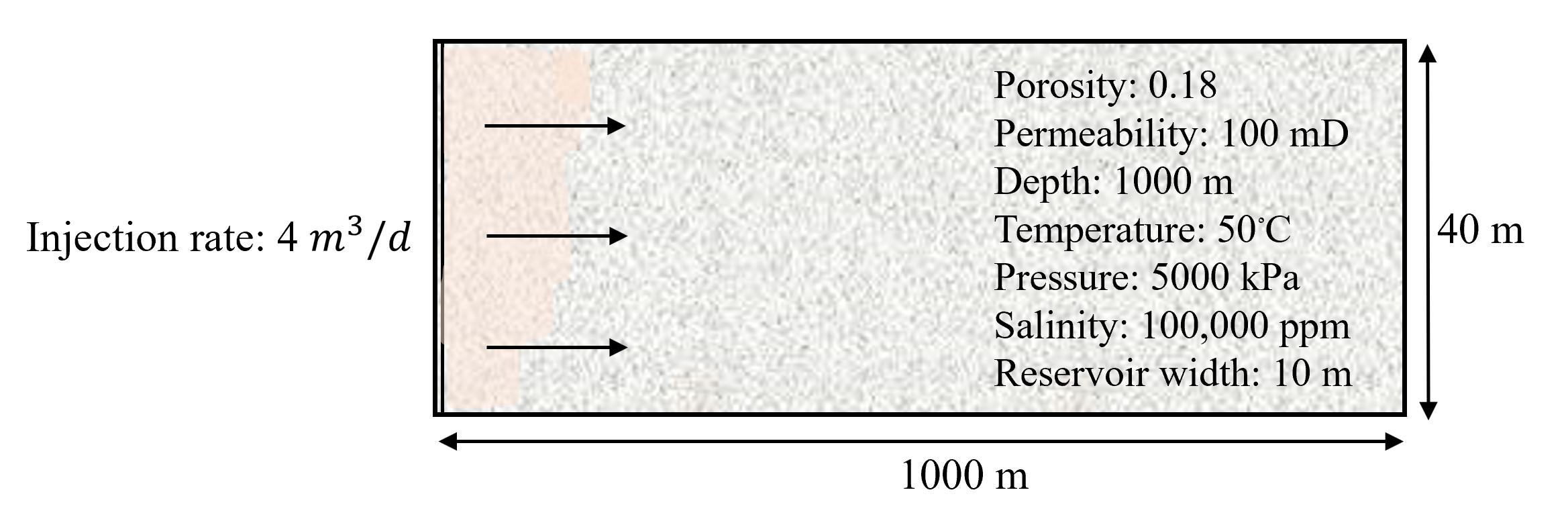}
    \caption{Schematic of a rectangular 1D flow field (Modified from Noh et al., \citep{noh2007implications}).}
    \label{fig:1DGas_reservoir}

\end{figure}

The training process is illustrated in Figure.\ref{fig:Gas_case}. It demonstrated that the PINN effectively handled two discontinuities, with the leading shock traveling significantly faster than the trailing shock. The final training loss achieved was 3.28E-07, and the error margin was 3.90\%.
The PINN-trained solution map was compared with the analytical solution in Figure. \ref{fig:2D_gas}, where warmer colors indicate areas of high gas saturation, and cooler colors signify zones of high water saturation. The gas saturations at the leading  leading ($S_{g 1}$) and the training shocks ($S_{g 2}$) were identified as 0.37 and 0.56, respectively. 
By calculating the derivative of the flux function at $S_{g 1}$ and $S_{g 2}$, we can determine the traveling velocities of the two gas fronts. Assuming CO\(_2\) is injected into a reservoir, as Figure.\ref{fig:1DGas_reservoir} shows, at a rate of 4 cubic meter per day (reservoir conditions) for 30 years, we can project the expansion of the CO\(_2\) plume. The distance the leading shock spreads is:
\begin{equation}
\left.x\right|_{S_{g 2}}=\left.\frac{d f(u)}{d u}\right|_{S_{g 2}} \times t_D \times L=985.4 \mathrm{~m}
\label{eq:gas11}
\end{equation}
The distance the trailing shock reaches is:
\begin{equation}
\left.x\right|_{S_{g 2}}=\left.\frac{d f(u)}{d u}\right|_{S_{g 2}} \times t_D \times L=69.7 \mathrm{~m}
\label{eq:gas22}
\end{equation}

\subsubsection{Dual-shock B-L model for purely gravitational flow}
Unlike the previous models where two shocks traveled in the same direction, this variant of the Buckley-Leverett (B-L) model introduces a scenario where two shocks induced by purely gravitational forces travel in opposite directions. This scenario poses a unique challenge, testing the adaptability and robustness of Physics-Informed Neural Networks (PINNs) in handling complex flow dynamics influenced by gravity. The viscosity and density ratios used for this model were sourced from Araujo et al. \citep{araujo2020numerical}: $\frac{\mu_w}{\mu_o}=0.25, \frac{\rho_w}{\rho_o}=1.25$. The initial conditions for water saturation were set as follows: 
\begin{equation}
s_w(z, 0)= \begin{cases}1-s_{o_r}, & z \leq 0 \\ s_{w_c}, & z>0\end{cases}
\end{equation}

\begin{figure}
    \centering
    \includegraphics[width=0.8\linewidth]{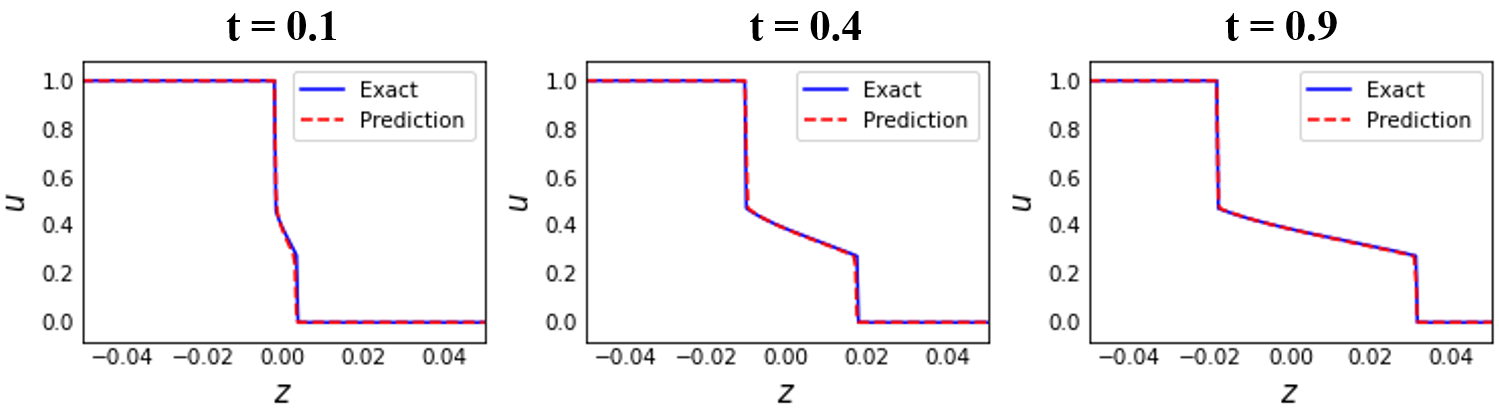}
    \caption{PINN solution for the model with two shocks traveling in the opposite direction.}
    \label{fig:Gravity_case}
\end{figure}

Figure.\ref{fig:Gravity_case} depicted the PINN solution versus distance at periodic time slices. At the midpoint ($z=0$), the saturation value remained constant. Due to gravity, one water shock and one oil shock formed and traveled at different speeds in opposite directions. 

\begin{figure}
    \centering
    \includegraphics[width=0.65\linewidth]{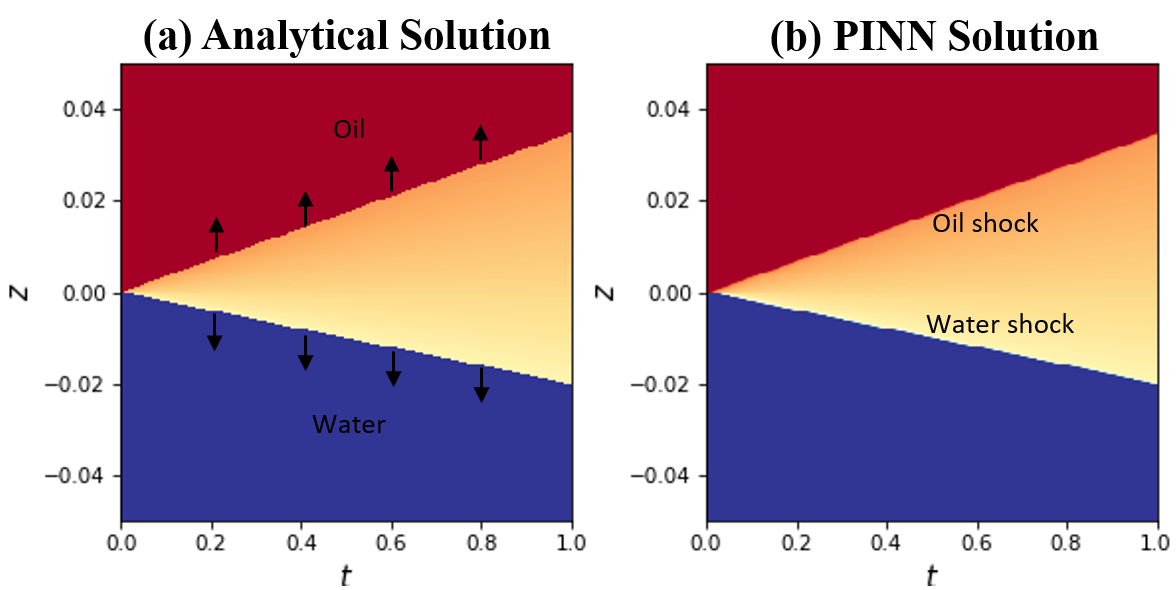}
    \caption{Analytical vs. PINN solution profiles: 2D comparison for the dual-shock B-L model for gravitational flow.}
    \label{fig:2D_gravity}
\end{figure}

The PINN trained solution was further compared with the analytical solution in Figure.\ref{fig:2D_gravity}. Oil shock, water shock, and the whole solution map were accurately captured by the PINN model.

\subsection{Inverse problems}
Within the framework of PINNs, inverse problems leverage observed (labeled) data to unravel the hidden parameters in the governing equations, such as rock and fluid properties in this study, thereby enabling the comprehensive prediction of the system's behavior over space and time. The training configurations for inverse PINNs mirror those of forward PINNs, including neural network architecture, initialization method, and optimization strategies. However, the composition of the loss function of inverse problems comprises solely the observed data error and the PDE residual error, as defined by Eq.\ref{eq:inverseloss}. Initial and boundary conditions are unknown in these scenarios.  

In addition to $L_2$-norm error and loss, the parameter error is employed to assess the performance of inverse PINN training, defined as:
\begin{equation}
\text{parameter error }=\frac{\mid \text{param}_{\text{true }}-\text{param}_{\text{estimated}}\mid^2}{\mid \text{param}_{\text{true}}\mid^2}
\end{equation}
It is important to point out that the fractional flow is not predefined but dynamically constructed during each iteration, with the front saturation calculated by Eq.\ref{eq:WC4}. This iterative refinement ensures that the PDE parameters and their constructions are continuously updated to align with observed data.

The forthcoming subsections will delve into the results of inverse training within the base scenario, which focuses on a single learnable parameter. We will examine how variations in the quantity and quality of sampling data, as well as the size of collocation points, impact PINN performance. Subsequently, we will address the complexities of learning with two parameters in the context of the Buckley-Leverett model.

\subsubsection{Base case}
For the base case, the focus was on learning the mobility ratio ($M$) , with gravity effects momentarily set aside. We assembled a dataset comprising 10,000 collocation points alongside 10,000 labeled data points, gathered through Latin Hypercube Sampling (LHS). This method ensured broad coverage across both time and spatial domains, aiming for thorough characterization of the system's dynamics. A learning rate of 1E-4 was employed for the neural network weights and a learning rate of 1E-3 was used for the learnable parameter. 

\begin{figure}
    \centering
    \includegraphics[width=0.6\linewidth]{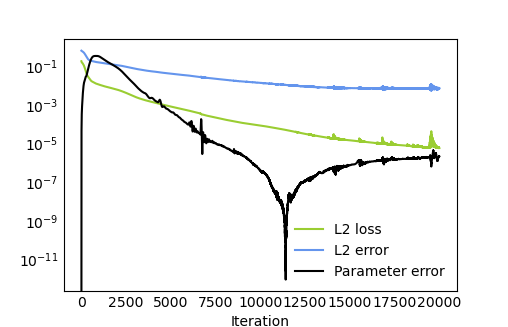}
    \caption{Inverse training of base case: evolution of loss, error, and parameter error.}
    \label{fig:Inverse_base}
\end{figure}

The progression of loss, error, and parameter error throughout the training process was depicted in Figure.\ref{fig:Inverse_base}. We identified the optimal model at an iteration where both the parameter error and solution error showed reductions compared to preceding values. Eventually, at iteration 11,416, the model precisely predicted $M$ to be 1.000000119 (true M=1), achieving an  $L_2$-norm error of 1.18\% and a loss of 6.17E-05. 
Expanding our analysis, we adjusted $M$ values for training and compiled the results in Table\ref{tab:inverse_cases}. For instance, the M1 case yielded an estimated M value of 0.09999999404 (true $M=0.1$) and for case M2, the NN model estimated $M$ to be 9.99998664856 (true $M=10$). The inverse PINN training undertaken in these cases consistently produced commendable results regarding parameter error, solution error, and the loss function.

To facilitate a more in-depth investigation on PINNs' performance with various sampling strategies of labeled data, eight additional cases were performed. The outcomes of these experiments, detailed in Table\ref{tab:inverse_cases}, were categorized into three groups for analysis.

\begin{table}
  \caption{Summary of Inverse PINN Training Cases (One Learnable Parameter)}
  \label{tab:inverse_cases}
  \centering
  \begin{tabular}{lccccc}
    \hline
    \textbf{Case} & \textbf{Col. Data} & \textbf{Lab. Data} & \textbf{Param Error} & \textbf{Error} & \textbf{Loss} \\
    \hline
    base       & 10k  & 10k  & 1.42E-14  & 0.01182  & 6.17E-05 \\
    M1 (M=0.1) & 10k  & 10k  & 3.56E-15  & 0.01428  & 9.08E-05 \\
    M2 (M=10)  & 10k  & 10k  & 1.78E-12  & 0.01846  & 8.66E-05 \\
    D1         & 10k  & 1k   & 6.96E-13  & 0.02306  & 2.23E-06 \\
    D2         & 10k  & 100  & 1.42E-14  & 0.03730  & 6.67E-06 \\
    D3         & 10k  & 10   & 3.55E-15  & 0.11830  & 4.73E-07 \\
    C1         & 10k  & 1\% noise & 8.88E-14  & 0.00997  & 1.41E-04 \\
    C2         & 10k  & 3\% noise & 1.28E-13  & 0.01333  & 9.63E-04 \\
    C3         & 10k  & 5\% noise & 4.30E-13  & 0.02476  & 2.74E-03 \\
    P1         & 1k   & 10k  & 6.96E-13  & 0.01325  & 4.63E-05 \\
    P2         & 100  & 10k  & 1.15E-09  & 0.01342  & 5.05E-05 \\
    \hline
  \end{tabular}
\end{table}

\subsubsection{Sensitivity analysis on sampling size}
\begin{figure}
    \centering
    \includegraphics[width=0.65\linewidth]{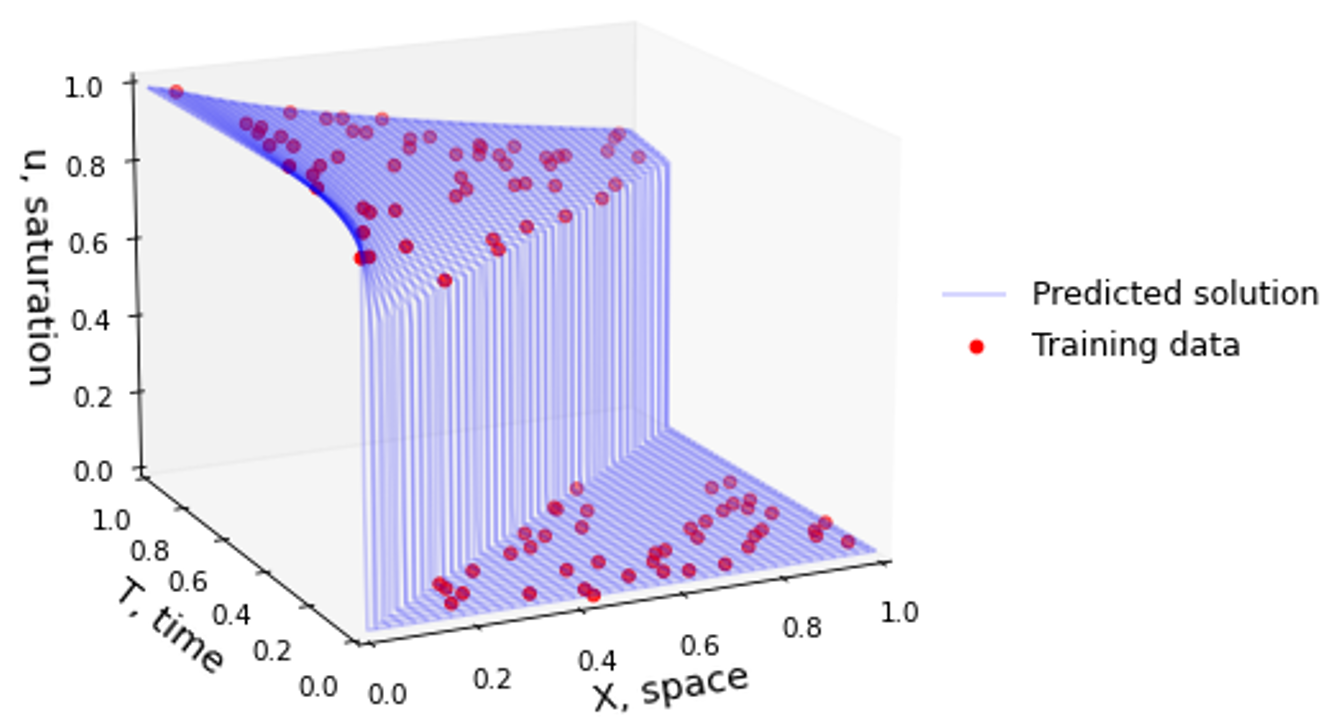}
    \caption{PINN solution map learnt from 100 labeled data.}
    \label{fig:100SamplePoints}
\end{figure}

\begin{figure}
    \centering
    \includegraphics[width=0.9\linewidth]{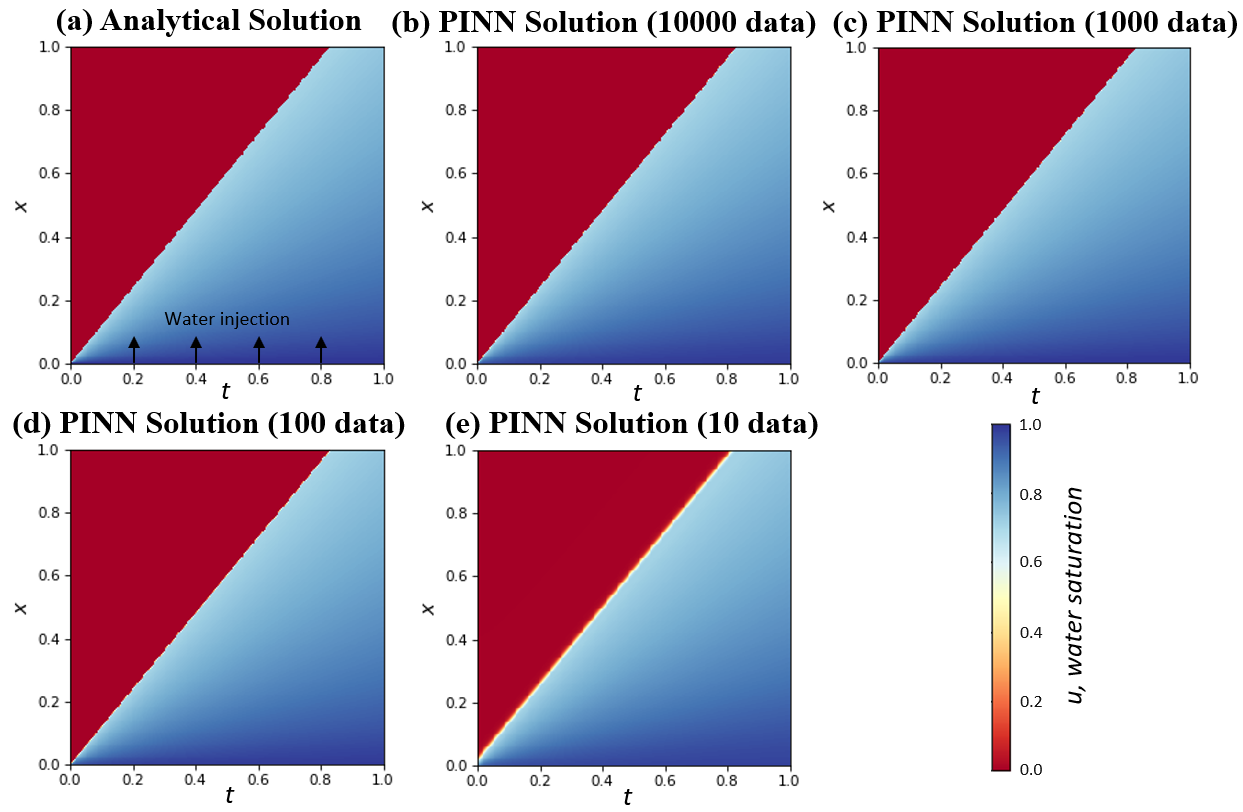}
    \caption{Analytical vs. PINN solution profiles: 2D comparison for various data sampling sizes.}
    \label{fig:inverse_errors_size}
\end{figure}

Ideally, the number of labeled data points should be comparable to the number of collocation points, such as the base, M1, and M2 cases. However, data scarcity in scientific and engineering contexts often necessitates an examination of how the effectiveness of PINNs fluctuates with varying sizes of labeled data. To address this, we conducted experiments to assess PINN performance with progressively smaller datasets in the D1, D2, and D3 cases, which utilized 1,000, 100, and 10 labeled data points, respectively. Figure.\ref{fig:100SamplePoints} illustrated the model's learning outcomes using just 100 labeled data points. 
The results from these cases, as presented in Table\ref{tab:inverse_cases}, validated that our method for selecting the optimal model consistently facilitated reliable parameter estimation across various datasets. Despite the reduced data sizes, the recorded loss values for these PINN models remained within modest ranges. However, as depicted in Figure.\ref{fig:inverse_errors_size},  the clarity in distinguishing between high and low saturation regions diminished with smaller sample sizes, indicating that solution prediction accuracy degraded as the number of sampling points decreased. Training with as few as 10 data points led to a significant decline in predictive performance, with the loss increasing by an order of magnitude compared to the base case.

\subsubsection{Sensitivity analysis on sampling purity}
\begin{figure}
    \centering
    \includegraphics[width=0.6\linewidth]{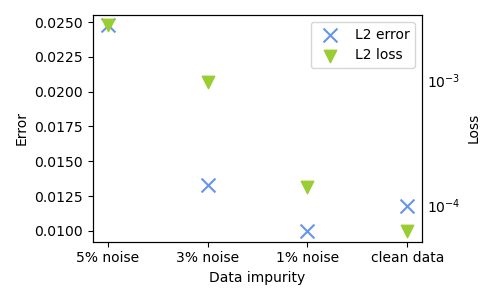}
    \caption{PINN prediction accuracy relative to data purity level.}
    \label{fig:inverse_errors_impurity}
\end{figure}

Beyond the impact of sampling size, given that real-world data collection frequently encounters the challenge of noise or impurities, we assessed the resilience of PINNs against such imperfections by introducing Gaussian noise at varying intensities of  1\% (C1 case), 3\% (C2 case), and 5\% (C3 case) into the pristine base case data. The corresponding training outcomes detailed in Table\ref{tab:inverse_cases}. Our results indicated that PINNs could successfully estimate the unknown parameter $M$ despite slight data corruption. For better visualization,  errors and losses of these four cases were plotted in Figure.\ref{fig:inverse_errors_impurity}. In the noise-free base case, the error and loss were recorded at 1.18\% and 6.17E-05, respectively. The introduction of 1\% noise slightly altered the results to an error of 1.00\% (lower than that of the base case, likely attributable to sampling variability) and a loss of 1.41E-04. At a 3\% noise level, the error and loss increased to  1.33\% and 9.63E-04, respectively. With 5\% noise level, the most corrupted scenario, the results further deteriorated to an error of  2.48\% and a loss of 2.74E-03.
Despite the apparent trend that increased corruption impaired PINN solution accuracy, the error and loss values, even at a 5\% noise level, remained relatively modest. This demonstrated the robustness of PINN training against minor noise interference, enhancing their applicability in real-world scenarios where data often comes with inherent inaccuracies.

\subsubsection{Sensitivity analysis on collocation data size}
\begin{figure}
    \centering
    \includegraphics[width=1.0\linewidth]{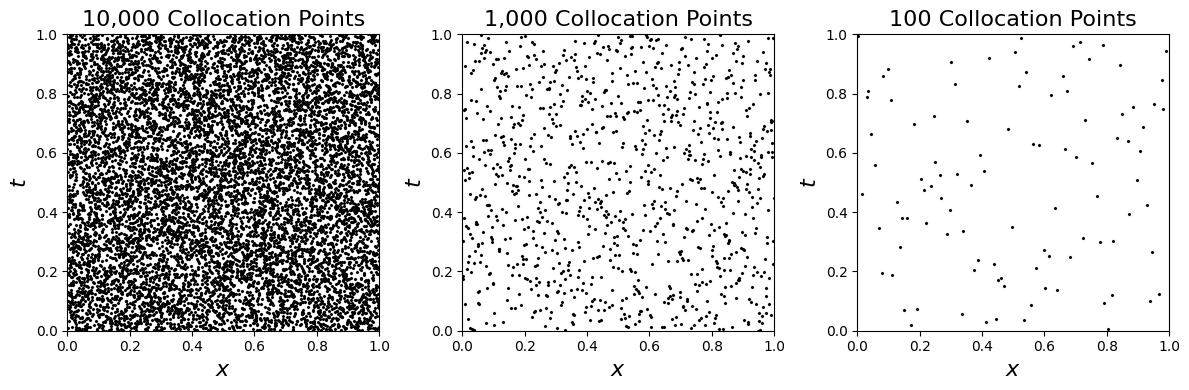}
    \caption{Comparison of collocation points for base case, P1 case, and P2 case.}
    \label{fig:collocationP}
\end{figure}

Unlike labeled data, collocation data generated from governing PDEs across spatial and temporal domains serve as 'free data' that ensure the neural network solution adheres to the underlying physical laws. Ideally, a large number and thorough distribution of collocation points should be used to guarantee the performance of PINNs, provided computational resources permit. To test the robustness of PINNs, we conducted two additional cases alongside the base case. The base case used 10,000 collocation points, while the additional cases, P1 and P2, used 1,000 and 100 collocation points, respectively. Figure.\ref{fig:collocationP} illustrates the different sizes of input collocation points for these cases.

The training errors, losses, and accuracy of parameter estimation for these cases were summarized in Table\ref{tab:inverse_cases}. The training losses remained relatively stable across the three cases. In the base case, which utilized 10,000 collocation points, the training error was found to be 1.18\%, with a corresponding estimated value of the mobility ratio of 1.00000011921. For the P1 case, which used 1,000 collocation points, the training error increased slightly to 1.33\%, with the estimated $M$ being 1.00000083447. In the P2 case, with only 100 collocation points, the training error further increased to 1.34\%, and the loss was 5.05E-05, with the estimated $M$ being 0.99996614456. A smaller size of collocation data points required more iterations of training to approach the exact solutions. However, a maximum iteration threshold of 20,000 was adequate for all three cases to achieve well-trained models. Additionally, the Latin Hypercube Sampling (LHS) method ensured a relatively uniform distribution of collocation data across space and time, contributing to the robustness of PINN performance. In summary, while there was a general trend of slightly lower accuracy with fewer collocation points, PINNs exhibited significant robustness in their performance.

\subsubsection{Two learnable parameters}
While vanilla PINNs demonstrate promising performance for inverse problems involving a single parameter, this subsection explores their capability to tackle the Buckley-Leverett model with two unknown parameters in the governing PDE: the mobility ratio ($M$) and the gravity term ($N \sin \alpha$). 10,000 labeled data points and 10,000 collocation points were used as inputs. 

The journey of training with two parameters proved to be considerably more complex, marked by notable fluctuations in losses and errors. To manage this complexity, three individual optimizers were deployed—one for the model hyperparameters and one each for the mobility ratio and gravity term—alongside careful adjustments of the learning rate for each optimizer. Consequently, the training duration extended to 65 minutes compared to approximately 15 minutes for the one-parameter case.

\begin{figure}
    \centering
    \includegraphics[width=0.7\linewidth]{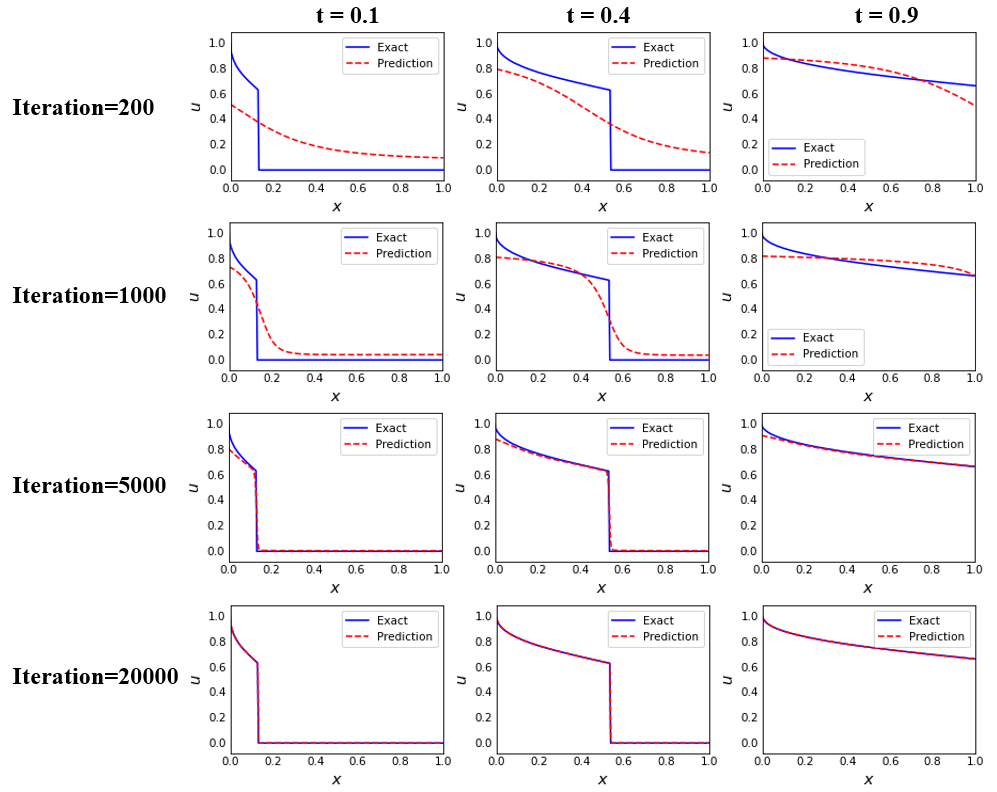}
    \caption{Evolution of solution profiles during inverse PINN training for the two-parameter case.}
    \label{fig:Inverse2_case}
\end{figure}

Through meticulous adjustments and optimization, stable convergence was achieved. The final error and loss were recorded at 1.43\% and 2.79E-6, respectively. The evolution of solution profiles during training  was displayed in Figure.\ref{fig:Inverse2_case}. The PINN estimated $M$ to be 0.999979854 (true value = 1) and the gravity term to be -1.00098896 (true value = -1), showcasing the model's high accuracy even with increased parameter complexity. 

\section{Conclusion and discussion}
\label{sec:Conclusion and discussion}
This research leverages state-of-the-art Physics-Informed Neural Network (PINN) techniques to solve and discover Buckley-Leverett equations and their variants, which exhibit intricate solution behaviors. Focusing on real-world petroleum engineering challenges, such as water flooding in subsurface hydrocarbon reservoirs and carbon sequestration in saline aquifers, we have identified several key findings:
\begin{enumerate}
\item The success of PINN training for the Buckley-Leverett model critically depends on Welge's construction of a convex hull for the original flux function. This method imposes precise physical constraints on the solutions, eliminating issues of multiple saturation values at a single location related to the original fractional flow. This strategy is generally applicable to other nonlinear hyperbolic PDEs exhibiting discontinuities.
\item PINNs in their most elemental forms efficiently solve the Buckley-Leverett equation across different fluid mobility ratios and gravity terms, without relying on labeled data. Our findings indicate that lower mobility ratios and upward-inclined reservoirs are favorable for delaying water breakthrough, thus enhancing oil recoveries in water-displacing-oil processes.
\item Vanilla PINNs demonstrate the capability to resolve not just a single saturation shock but also dual shocks in both a semi-miscible gas-displacing-water process, where shocks travel in the same direction, and in purely gravitational flow processes, where shocks travel in opposite directions. The presence of additional discontinuities, due to inter-phase solubility and dominant gravity effects, does not impede the effectiveness of PINNs. This capability provides a valuable tool for modeling the spread of a CO\(_2\) plume and gravitational flows.
    \item Utilizing observed data, inverse PINNs precisely identify hidden parameters in the governing equations, such as the mobility ratio. The constraints imposed by the governing equations reduce the dependence of inverse PINNs on labeled data. Our sensitivity analysis shows that PINNs demonstrate resilience to data impurities of up to 5\% and cope well with moderate data shortages. Furthermore, PINNs exhibit significant robustness when varying the number of collocation points, maintaining accuracy even with reduced data sizes. 
\item Inverse PINNs are capable of identifying multiple parameters in the Buckley-Leverett equation, enabling comprehensive mapping of the entire solution space through meticulous adjustments of learning rates for individual optimizers.

\end{enumerate}

PINNs exhibit independence from labeled data and excel in extrapolation or prediction capabilities for forward problems, outperforming other machine learning methods. Compared to numerical methods, the meshless characteristic of PINNs eliminates the need for fine grid blocks to track shock front movements, thereby avoiding the discretization errors associated with numerical simulation. However, it is essential to recognize that PINNs are not designed to replace but to augment traditional simulation methods. Currently, PINN techniques are in their developmental stages and encounter challenges in modeling complex physical phenomena, similar to their governing models. For instance, the Buckley-Leverett equation assumes constant rock and fluid properties and neglects capillary pressure, which restricts the generalizability of PINNs in highly heterogeneous and fractured reservoirs exhibiting hysteresis. Future research efforts will focus on overcoming these challenges and extending the application of PINNs to two-dimensional and even three-dimensional models, as well as incorporating capillarity, thereby broadening their impact in engineering and scientific research fields.

\section{Abbreviations}
\begin{description}
    \item[ML] Machine Learning
    \item[PIML] Physics-Informed Machine Learning
    \item[CCUS] Carbon Capture, Utilization, and Storage
    \item[PINN] Physics-Informed Neural Network
    \item[PDE]  Partial Differential Equation
    \item[BL] Buckley-Leverett
    \item[NN] Neural Network
    \item[AD] Automation Differentiation
    \item[LHS] Latin Hypercube Sampling
\end{description}

\bibliographystyle{apalike}
\bibliography{ref}  

\providecommand{\latin}[1]{#1}
\makeatletter
\providecommand{\doi}
  {\begingroup\let\do\@makeother\dospecials
  \catcode`\{=1 \catcode`\}=2 \doi@aux}
\providecommand{\doi@aux}[1]{\endgroup\texttt{#1}}
\makeatother
\providecommand*\mcitethebibliography{\thebibliography}
\csname @ifundefined\endcsname{endmcitethebibliography}
  {\let\endmcitethebibliography\endthebibliography}{}
\begin{mcitethebibliography}{27}
\providecommand*\natexlab[1]{#1}
\providecommand*\mciteSetBstSublistMode[1]{}
\providecommand*\mciteSetBstMaxWidthForm[2]{}
\providecommand*\mciteBstWouldAddEndPuncttrue
  {\def\EndOfBibitem{\unskip.}}
\providecommand*\mciteBstWouldAddEndPunctfalse
  {\let\EndOfBibitem\relax}
\providecommand*\mciteSetBstMidEndSepPunct[3]{}
\providecommand*\mciteSetBstSublistLabelBeginEnd[3]{}
\providecommand*\EndOfBibitem{}
\mciteSetBstSublistMode{f}
\mciteSetBstMaxWidthForm{subitem}{(\alph{mcitesubitemcount})}
\mciteSetBstSublistLabelBeginEnd
  {\mcitemaxwidthsubitemform\space}
  {\relax}
  {\relax}

\bibitem[Latrach \latin{et~al.}(2023)Latrach, Malki, Morales, Mehana, and
  Rabiei]{latrach2023critical}
Latrach,~A.; Malki,~M.~L.; Morales,~M.; Mehana,~M.; Rabiei,~M. A Critical
  Review of Physics-Informed Machine Learning Applications in Subsurface Energy
  Systems. \emph{arXiv preprint arXiv:2308.04457} \textbf{2023}, \relax
\mciteBstWouldAddEndPunctfalse
\mciteSetBstMidEndSepPunct{\mcitedefaultmidpunct}
{}{\mcitedefaultseppunct}\relax
\EndOfBibitem
\bibitem[Kesireddy \latin{et~al.}(2023)Kesireddy, Kompantsev, Dey, Gildin,
  Losoya, and Vishnumolakala]{kesireddy2023maximizing}
Kesireddy,~V.; Kompantsev,~G.; Dey,~S.; Gildin,~E.; Losoya,~E.~Z.;
  Vishnumolakala,~N. Maximizing Efficiency of Deep-Reinforcement Learning
  Agents in Autonomous Directional Drilling with Hyperparameter Optimization.
  SPE/AAPG/SEG Unconventional Resources Technology Conference. 2023; p
  D031S063R004\relax
\mciteBstWouldAddEndPuncttrue
\mciteSetBstMidEndSepPunct{\mcitedefaultmidpunct}
{\mcitedefaultendpunct}{\mcitedefaultseppunct}\relax
\EndOfBibitem
\bibitem[Badawi and Gildin(2023)Badawi, and Gildin]{badawi2023physics}
Badawi,~D.; Gildin,~E. Physics-Informed Neural Network for the Transient
  Diffusivity Equation in Reservoir Engineering. \emph{arXiv preprint
  arXiv:2309.17345} \textbf{2023}, \relax
\mciteBstWouldAddEndPunctfalse
\mciteSetBstMidEndSepPunct{\mcitedefaultmidpunct}
{}{\mcitedefaultseppunct}\relax
\EndOfBibitem
\bibitem[Wang and Chen(2023)Wang, and Chen]{wang2023insights}
Wang,~H.; Chen,~S. Insights into the Application of Machine Learning in
  Reservoir Engineering: Current Developments and Future Trends.
  \emph{Energies} \textbf{2023}, \emph{16}, 1392\relax
\mciteBstWouldAddEndPuncttrue
\mciteSetBstMidEndSepPunct{\mcitedefaultmidpunct}
{\mcitedefaultendpunct}{\mcitedefaultseppunct}\relax
\EndOfBibitem
\bibitem[Raissi \latin{et~al.}(2019)Raissi, Perdikaris, and
  Karniadakis]{raissi2019physics}
Raissi,~M.; Perdikaris,~P.; Karniadakis,~G.~E. Physics-informed neural
  networks: A deep learning framework for solving forward and inverse problems
  involving nonlinear partial differential equations. \emph{Journal of
  Computational physics} \textbf{2019}, \emph{378}, 686--707\relax
\mciteBstWouldAddEndPuncttrue
\mciteSetBstMidEndSepPunct{\mcitedefaultmidpunct}
{\mcitedefaultendpunct}{\mcitedefaultseppunct}\relax
\EndOfBibitem
\bibitem[Fraces \latin{et~al.}(2020)Fraces, Papaioannou, and
  Tchelepi]{fraces2020physics}
Fraces,~C.~G.; Papaioannou,~A.; Tchelepi,~H. Physics informed deep learning for
  transport in porous media. Buckley Leverett problem. \emph{arXiv preprint
  arXiv:2001.05172} \textbf{2020}, \relax
\mciteBstWouldAddEndPunctfalse
\mciteSetBstMidEndSepPunct{\mcitedefaultmidpunct}
{}{\mcitedefaultseppunct}\relax
\EndOfBibitem
\bibitem[Buckley and Leverett(1942)Buckley, and Leverett]{buckley1942mechanism}
Buckley,~S.~E.; Leverett,~M. Mechanism of fluid displacement in sands.
  \emph{Transactions of the AIME} \textbf{1942}, \emph{146}, 107--116\relax
\mciteBstWouldAddEndPuncttrue
\mciteSetBstMidEndSepPunct{\mcitedefaultmidpunct}
{\mcitedefaultendpunct}{\mcitedefaultseppunct}\relax
\EndOfBibitem
\bibitem[Fuks and Tchelepi(2020)Fuks, and Tchelepi]{fuks2020limitations}
Fuks,~O.; Tchelepi,~H.~A. Limitations of physics informed machine learning for
  nonlinear two-phase transport in porous media. \emph{Journal of Machine
  Learning for Modeling and Computing} \textbf{2020}, \emph{1}\relax
\mciteBstWouldAddEndPuncttrue
\mciteSetBstMidEndSepPunct{\mcitedefaultmidpunct}
{\mcitedefaultendpunct}{\mcitedefaultseppunct}\relax
\EndOfBibitem
\bibitem[Coutinho \latin{et~al.}(2023)Coutinho, Dall'Aqua, McClenny, Zhong,
  Braga-Neto, and Gildin]{coutinho2023physics}
Coutinho,~E. J.~R.; Dall'Aqua,~M.; McClenny,~L.; Zhong,~M.; Braga-Neto,~U.;
  Gildin,~E. Physics-informed neural networks with adaptive localized
  artificial viscosity. \emph{Journal of Computational Physics} \textbf{2023},
  112265\relax
\mciteBstWouldAddEndPuncttrue
\mciteSetBstMidEndSepPunct{\mcitedefaultmidpunct}
{\mcitedefaultendpunct}{\mcitedefaultseppunct}\relax
\EndOfBibitem
\bibitem[Rodriguez-Torrado \latin{et~al.}(2022)Rodriguez-Torrado, Ruiz,
  Cueto-Felgueroso, Green, Friesen, Matringe, and
  Togelius]{rodriguez2022physics}
Rodriguez-Torrado,~R.; Ruiz,~P.; Cueto-Felgueroso,~L.; Green,~M.~C.;
  Friesen,~T.; Matringe,~S.; Togelius,~J. Physics-informed attention-based
  neural network for hyperbolic partial differential equations: application to
  the Buckley--Leverett problem. \emph{Scientific reports} \textbf{2022},
  \emph{12}, 7557\relax
\mciteBstWouldAddEndPuncttrue
\mciteSetBstMidEndSepPunct{\mcitedefaultmidpunct}
{\mcitedefaultendpunct}{\mcitedefaultseppunct}\relax
\EndOfBibitem
\bibitem[Diab \latin{et~al.}(2022)Diab, Chaabi, Zhang, Arif, Alkobaisi, and
  Al~Kobaisi]{diab2022data}
Diab,~W.; Chaabi,~O.; Zhang,~W.; Arif,~M.; Alkobaisi,~S.; Al~Kobaisi,~M.
  Data-Free and Data-Efficient Physics-Informed Neural Network Approaches to
  Solve the Buckley--Leverett Problem. \emph{Energies} \textbf{2022},
  \emph{15}, 7864\relax
\mciteBstWouldAddEndPuncttrue
\mciteSetBstMidEndSepPunct{\mcitedefaultmidpunct}
{\mcitedefaultendpunct}{\mcitedefaultseppunct}\relax
\EndOfBibitem
\bibitem[Welge(1952)]{welge1952simplified}
Welge,~H.~J. A simplified method for computing oil recovery by gas or water
  drive. \emph{Journal of Petroleum Technology} \textbf{1952}, \emph{4},
  91--98\relax
\mciteBstWouldAddEndPuncttrue
\mciteSetBstMidEndSepPunct{\mcitedefaultmidpunct}
{\mcitedefaultendpunct}{\mcitedefaultseppunct}\relax
\EndOfBibitem
\bibitem[Magzymov \latin{et~al.}(2021)Magzymov, Ratnakar, Dindoruk, and
  Johns]{magzymov2021evaluation}
Magzymov,~D.; Ratnakar,~R.~R.; Dindoruk,~B.; Johns,~R.~T. Evaluation of machine
  learning methodologies using simple physics based conceptual models for flow
  in porous media. SPE Annual Technical Conference and Exhibition? 2021; p
  D021S038R004\relax
\mciteBstWouldAddEndPuncttrue
\mciteSetBstMidEndSepPunct{\mcitedefaultmidpunct}
{\mcitedefaultendpunct}{\mcitedefaultseppunct}\relax
\EndOfBibitem
\bibitem[Fraces and Tchelepi(2021)Fraces, and Tchelepi]{fraces2021physics}
Fraces,~C.~G.; Tchelepi,~H. Physics informed deep learning for flow and
  transport in porous media. SPE Reservoir Simulation Conference? 2021; p
  D011S006R002\relax
\mciteBstWouldAddEndPuncttrue
\mciteSetBstMidEndSepPunct{\mcitedefaultmidpunct}
{\mcitedefaultendpunct}{\mcitedefaultseppunct}\relax
\EndOfBibitem
\bibitem[Cuomo \latin{et~al.}(2022)Cuomo, Di~Cola, Giampaolo, Rozza, Raissi,
  and Piccialli]{cuomo2022scientific}
Cuomo,~S.; Di~Cola,~V.~S.; Giampaolo,~F.; Rozza,~G.; Raissi,~M.; Piccialli,~F.
  Scientific machine learning through physics--informed neural networks: Where
  we are and what’s next. \emph{Journal of Scientific Computing}
  \textbf{2022}, \emph{92}, 88\relax
\mciteBstWouldAddEndPuncttrue
\mciteSetBstMidEndSepPunct{\mcitedefaultmidpunct}
{\mcitedefaultendpunct}{\mcitedefaultseppunct}\relax
\EndOfBibitem
\bibitem[Brooks and Corey(1966)Brooks, and Corey]{brooks1966properties}
Brooks,~R.~H.; Corey,~A.~T. Properties of porous media affecting fluid flow.
  \emph{Journal of the irrigation and drainage division} \textbf{1966},
  \emph{92}, 61--88\relax
\mciteBstWouldAddEndPuncttrue
\mciteSetBstMidEndSepPunct{\mcitedefaultmidpunct}
{\mcitedefaultendpunct}{\mcitedefaultseppunct}\relax
\EndOfBibitem
\bibitem[Araujo \latin{et~al.}(2020)Araujo, Rodr{\'\i}guez-Berm{\'u}dez, and
  Rodr{\'\i}guez-N{\'u}{\~n}ez]{araujo2020numerical}
Araujo,~I.~L.; Rodr{\'\i}guez-Berm{\'u}dez,~P.;
  Rodr{\'\i}guez-N{\'u}{\~n}ez,~Y. Numerical study for two-phase flow with
  gravity in homogeneous and piecewise-homogeneous porous media. \emph{TEMA
  (S{\~a}o Carlos)} \textbf{2020}, \emph{21}, 21--41\relax
\mciteBstWouldAddEndPuncttrue
\mciteSetBstMidEndSepPunct{\mcitedefaultmidpunct}
{\mcitedefaultendpunct}{\mcitedefaultseppunct}\relax
\EndOfBibitem
\bibitem[Dake(1983)]{dake1983fundamentals}
Dake,~L.~P. \emph{Fundamentals of reservoir engineering}; Elsevier, 1983\relax
\mciteBstWouldAddEndPuncttrue
\mciteSetBstMidEndSepPunct{\mcitedefaultmidpunct}
{\mcitedefaultendpunct}{\mcitedefaultseppunct}\relax
\EndOfBibitem
\bibitem[LeVeque and Leveque(1992)LeVeque, and Leveque]{leveque1992numerical}
LeVeque,~R.~J.; Leveque,~R.~J. \emph{Numerical methods for conservation laws};
  Springer, 1992; Vol. 214\relax
\mciteBstWouldAddEndPuncttrue
\mciteSetBstMidEndSepPunct{\mcitedefaultmidpunct}
{\mcitedefaultendpunct}{\mcitedefaultseppunct}\relax
\EndOfBibitem
\bibitem[Green \latin{et~al.}(1998)Green, Willhite, \latin{et~al.}
  others]{green1998enhanced}
Green,~D.~W.; Willhite,~G.~P., \latin{et~al.}  \emph{Enhanced oil recovery};
  Henry L. Doherty Memorial Fund of AIME, Society of Petroleum Engineers~…,
  1998; Vol.~6\relax
\mciteBstWouldAddEndPuncttrue
\mciteSetBstMidEndSepPunct{\mcitedefaultmidpunct}
{\mcitedefaultendpunct}{\mcitedefaultseppunct}\relax
\EndOfBibitem
\bibitem[Li \latin{et~al.}(2024)Li, Jia, Yao, Sepehrnoori, Abushaikha, and
  Liu]{li2024investigation}
Li,~L.; Jia,~C.; Yao,~J.; Sepehrnoori,~K.; Abushaikha,~A.; Liu,~Y. An
  Investigation of Gas-Fingering Behavior during CO2 Flooding in Acid
  Stimulation Formations. \emph{SPE Journal} \textbf{2024}, 1--18\relax
\mciteBstWouldAddEndPuncttrue
\mciteSetBstMidEndSepPunct{\mcitedefaultmidpunct}
{\mcitedefaultendpunct}{\mcitedefaultseppunct}\relax
\EndOfBibitem
\bibitem[Noh \latin{et~al.}(2007)Noh, Lake, Bryant, and
  Araque-Martinez]{noh2007implications}
Noh,~M.; Lake,~L.~W.; Bryant,~S.~L.; Araque-Martinez,~A. Implications of
  coupling fractional flow and geochemistry for CO2 injection in aquifers.
  \emph{SPE Reservoir Evaluation \& Engineering} \textbf{2007}, \emph{10},
  406--414\relax
\mciteBstWouldAddEndPuncttrue
\mciteSetBstMidEndSepPunct{\mcitedefaultmidpunct}
{\mcitedefaultendpunct}{\mcitedefaultseppunct}\relax
\EndOfBibitem
\bibitem[Burton \latin{et~al.}(2009)Burton, Kumar, and Bryant]{burton2009co2}
Burton,~M.; Kumar,~N.; Bryant,~S.~L. CO2 injectivity into brine aquifers: Why
  relative permeability matters as much as absolute permeability. \emph{Energy
  Procedia} \textbf{2009}, \emph{1}, 3091--3098\relax
\mciteBstWouldAddEndPuncttrue
\mciteSetBstMidEndSepPunct{\mcitedefaultmidpunct}
{\mcitedefaultendpunct}{\mcitedefaultseppunct}\relax
\EndOfBibitem
\bibitem[Azizi and Cinar(2013)Azizi, and Cinar]{azizi2013approximate}
Azizi,~E.; Cinar,~Y. Approximate analytical solutions for CO2 injectivity into
  saline formations. \emph{SPE Reservoir Evaluation \& Engineering}
  \textbf{2013}, \emph{16}, 123--133\relax
\mciteBstWouldAddEndPuncttrue
\mciteSetBstMidEndSepPunct{\mcitedefaultmidpunct}
{\mcitedefaultendpunct}{\mcitedefaultseppunct}\relax
\EndOfBibitem
\bibitem[Bai \latin{et~al.}(2024)Bai, Jia, Hu, Alsousy, Lu, and
  Sepehrnoori]{bai2024storage}
Bai,~F.; Jia,~C.; Hu,~J.; Alsousy,~A.; Lu,~Y.; Sepehrnoori,~K. Storage capacity
  comparison of hydrogen and carbon dioxide in heterogeneous aquifers.
  \emph{Gas Science and Engineering} \textbf{2024}, \emph{121}, 205182\relax
\mciteBstWouldAddEndPuncttrue
\mciteSetBstMidEndSepPunct{\mcitedefaultmidpunct}
{\mcitedefaultendpunct}{\mcitedefaultseppunct}\relax
\EndOfBibitem
\bibitem[Ketcheson \latin{et~al.}(2020)Ketcheson, LeVeque, and
  Del~Razo]{ketcheson2020riemann}
Ketcheson,~D.~I.; LeVeque,~R.~J.; Del~Razo,~M.~J. \emph{Riemann problems and
  Jupyter solutions}; SIAM, 2020; Vol.~16\relax
\mciteBstWouldAddEndPuncttrue
\mciteSetBstMidEndSepPunct{\mcitedefaultmidpunct}
{\mcitedefaultendpunct}{\mcitedefaultseppunct}\relax
\EndOfBibitem
\end{mcitethebibliography}

\end{document}